\documentclass[times,final]{elsarticle}

\usepackage{jcomp}
\usepackage{framed,multirow}

\usepackage{lineno,hyperref}
\usepackage[utf8]{inputenc}
\usepackage[version=4]{mhchem}
\usepackage{siunitx}
\usepackage{graphicx}
\usepackage{epstopdf, epsfig}
\usepackage{lipsum}
\usepackage{caption}
\usepackage{subcaption}
\usepackage{amsmath}
\usepackage{commath}
\usepackage{wrapfig}
\usepackage{xcolor}
\usepackage{subfiles}
\usepackage{pdfpages}
\usepackage{verbatim}
\usepackage{multirow}
\usepackage{booktabs}
\usepackage{mathabx}
\usepackage{tikz}
\usepackage{bm}
\usetikzlibrary{shapes}

\journal{Journal of Computational Physics}

\biboptions{authoryear}

\graphicspath{
{./figures/}
{./figures/BurgersNEW/}
{./figures/SOD_NEW/}
{./figures/channel/}
}

\newcommand{\review}[1]{{\color{black} #1}}

\DeclareRobustCommand\sampleline[1]{%
  \tikz\draw[#1] (0,0) (0,\the\dimexpr\fontdimen22\textfont2\relax)
  -- (2em,\the\dimexpr\fontdimen22\textfont2\relax);%
}

\newcommand{\be}{\begin{equation}}
\newcommand{\ee}{\end{equation}}

\newcommand{\modelFullName}{Quasi-Spectral Viscosity}
\newcommand{\modelAcronym}{QSV}

\begin{document}

\begin{frontmatter}

\title{A unified \modelFullName\ (\modelAcronym) approach to shock capturing and large-eddy simulation
}
\author[aff1]{Victor C. B. Sousa\corref{mycorrespondingauthor}}
\cortext[mycorrespondingauthor]{Corresponding author}
\ead{vsousa@purdue.edu}

\verso{V. Sousa and C. Scalo}

\author[aff1,aff2]{Carlo Scalo}

\address[aff1]{School of Mechanical Engineering, Purdue University, West Lafayette, IN, 47907}
\address[aff2]{School of Aeronautics and Astronautics, Purdue University, West Lafayette, IN, 47907}

\received{23 March 2021}

\begin{abstract}

 \review{The \modelFullName\ (\modelAcronym) method is a novel closure for a high-order finite-difference discretization of the filtered compressible Navier-Stokes equations capable of unifying dynamic sub-filter-scale (SFS) modeling and shock capturing under a single mathematical framework. Its innovation lies in the introduction of a physical-space implementation of a spectral-like SFS dissipation term by leveraging residuals of filter operations, achieving two goals: (1) estimating the energy of the resolved solution near the grid cutoff; (2) imposing a plateau-cusp shape to the spectral distribution of the added dissipation. The \modelAcronym~approach has been tested in a variety of flows to showcase its capability to act interchangeably as: a shock capturing method, in the Shu-Osher, shock/vortex or shock/wall interactions problems; or as a SFS closure, in subsonic Taylor Green Vortex (TGV), and supersonic/hypersonic turbulent channel flows. \modelAcronym~ performs well compared to previous eddy-viscosity closures and shock capturing methods in such test cases. In a supersonic TGV flow, a case which exhibits shock/turbulence interactions, \modelAcronym~alone outperforms the simple superposition of separate numerical treatments for SFS turbulence and shocks. \modelAcronym's combined capability of simulating shocks and turbulence independently, as well as simultaneously, effectively achieves the unification of shock capturing and Large-Eddy Simulation.}

\end{abstract}


\end{frontmatter}




\section{Introduction} 
\label{sec:IntroLES}

Modeling approaches for hydrodynamic turbulence and shock formation have taken historically different paths, despite both phenomena being characterized by an energy cascade from large to small scales due to nonlinear interactions \citep{frisch1995turbulence,GuptaScalo_PRE_2018}. \review{This suggests that the modeling approach for such distinct phenomena could be indeed unified, allowing accurate and numerically stable results of highly compressible turbulent flows on relatively coarse grids with a single artificial-dissipation approach. An application that benefits from such development is the modeling of transitional or fully turbulent hypersonic boundary layers characterized by steep, shock-like, flow gradients that can arise as a result of both nonlinear waves or hydrodynamics. }

The Large-Eddy Simulation (LES) methodology was developed to relax the Reynolds number constraint on numerical simulations of hydrodynamic turbulence. In an LES, the original Navier-Stokes equations are filtered via a low-pass band filtering operation which commutes with the spatial and temporal derivatives in an attempt to separate the large, or filtered, scales from the small, sub filter scales (SFS). Although such an operation is able to seamlessly separate the scales when applied to linear terms, when it acts upon nonlinear components, an unclosed term, connected to energy flux between scales, appears. The genesis of such a term is connected to the dynamics of large and smalls scales being coupled. Ultimately, the LES method directly computes the evolution of the large scales by modeling the energy flow towards the small scales via a dissipative term. 

The effort spent in the development of subfilter-models for LES has been considerable and it was initially focused on incompressible flows. Most of these models are of the eddy-viscosity type, where the energy flux to the small scales is modeled as a viscous dissipation process within a fluid. In such models, the fluid's original kinematic viscosity field is augmented by an eddy-viscosity term ($\nu_t = \upsilon_c \ell$) whose magnitude is connected to a velocity scale near the filter's cutoff ($\upsilon_c$) and a mixing length ($\ell$). These models have been defined both in the physical and spectral space. 

The first SFS model was proposed by \citet{Smagorinsky_MWR_1963} who assumed that the small scale turbulent kinetic energy (TKE) production and dissipation are in equilibrium and that turbulence is in a state of statistical isotropy. One drawback of this model is that the SFS dissipation is always active: in regions of transitional flow or near boundaries, the model over predicts the dissipation leading to inaccurate results. This drawback was overcome by the introduction of the dynamic procedure, which instantaneously modulated the local dissipation by accessing information near the grid cutoff through test filtering the resolved scales and averaging over homogeneous directions  \citep{GermanoPMC91,lilly1992proposed}, or, more locally, over Lagrangian fluid particle paths \citep{meneveau1996lagrangian}. 

These models were later extended to the compressible Navier-Stokes equations. One example is \citet{moin1991dynamic}, who used the Favre filtered, continuity, momentum and internal energy equations to implement Germano's dynamic procedure for compressible flows. He proposed models for the SFS stresses that arise in the momentum equations and for the SFS internal energy transport while opting to neglect the contribution of the pressure-dilation and turbulence dissipation rate terms. Their LES approach showed good agreement against experiments and Direct Numerical Simulation on setups with turbulent Mach numbers up to $M_t = 0.4$. Moreover, as opposed to incompressible flows, where the trace of the SFS stress tensor is absorbed in the pressure term, this approach modeled it separately, following \citet{Yoshizawa_PoF_1986}'s parametrization. Although this model was applied in simulations with no shocks, {\it a posteriori} results reported that the contribution of the trace could be as high as 50\% of its deviatoric part.

These models, however, yield a flat wavenumber spectrum of SFS dissipation, which is inconsistent with the studies performed by \citet{kraichnan1976eddy}. He introduced the concept of a wavenumber-dependent eddy viscosity to model the energy transfer across a filter cutoff. \review{The motivation behind this choice lies in the fact that, if the primary filter cutoff lies in the inertial subrange, the resulting eddy viscosity is not a flat function of the resolved wavenumber space: it exhibits, rather, a plateau at low wavenumbers and a sharp rise near the grid cutoff, referred to as a \emph{plateau-cusp behavior}. }\citet{chollet1981parameterization} reached similar conclusions by adopting the eddy-damped quasi-normal Markovian theory (EDQNM). Models that fail to modulate the dissipation rate, without enhancing it near the cutoff wavenumber, lead to a spurious accumulation of energy at the smallest resolved scales, i.e. high wavenumber energy build up, as reported by \citet{moin1991dynamic}, for example. 

\citet{chollet1981parameterization} proposed an exponential fit to the plateau-cusp behavior in the spectral space with magnitude depending on a dimensional term comprising a velocity scale multiplied by a length scale. The information on such flow scales were extracted from the kinetic energy at the cutoff, $\sqrt{E_{k_c}/k_c}$. This model, \review{referred to as} spectral eddy viscosity (SEV), is also dynamic since at the early stages of the flow evolution, when there is still no energy at the cutoff, no SFS dissipation is introduced. Despite the solid theoretical foundations of this work, its applicability has been limited to homogeneous flows where the Navier-Stokes equations could be solved conveniently in Fourier spectral space. To extend this approach to inhomogeneous flows solved in the physical space, \citet{metais1992spectral} introduced the Structure Function (SF) model. In this approach a constant wavenumber viscosity spectrum whose value was connected to the wavenumber-averaged ``plateau-cusp'' theoretical viscosity curve and estimated the energy at the cutoff via a structure function. Although very good agreement with DNS is reported, high wavenumber spectral energy build-up was observed. 

The SF model was also used in the context of compressible LES simulations \citep{Normand_TCFD_1992,Ducros_TSF_1995} where the SFS internal energy transport was modeled via a constant turbulent Prandtl number assumption. Other assumptions needed to extend the Structure Function model to a compressible flow involve considering coherent large structures to be sufficiently separated from the isotropic SFS field and assumed to be not affected by compressibility. It was also mentioned that this would have no validity in the neighborhood of a shock although it could help in its numerical capturing. This model was used by \citet{Ducros_TSF_1995} to simulate temporally developing supersonic boundary layers where the LES model was shown to stabilize the calculation once turbulence has been fully developed with only small effects on the dynamics of the transitional waves.

\review{Just as in hydrodynamic turbulence, shocks in compressible flows arise from nonlinear wave steepening, entailing the generation of small scales \citep{GuptaScalo_PRE_2018}. This implies that the same mathematical structure of the filtered equations developed by the LES community could also be used, in principle, to model shock discontinuities. In spite of this, the numerical modeling of the two phenomena has historically followed two different paths.} For example, \citet{Ducros_JCP_1999} developed a shock capturing technique consisting of a sensor that triggers artificial dissipation in the shock region. The scheme was shown to thicken shocks so that they could be numerically resolved but the dissipation applied in the shock affected the turbulence that was interacting with it. Similar models of local artificial diffusivity (LAD) were developed for shock turbulence interactions for example, in high-order finite-difference simulations \citep{Cook_PoF_2007,Kawai_JCP_2008,Kawai_JCP_2010}, in an unstructured spectral difference framework \citep{Premasuthan_2014_I,Premasuthan_2014_II} and in flux-reconstruction schemes \citep{HagaKawai_JCP_2019}.

 Concomitantly, a different method called Spectral Vanishing Viscosity (SVV) arose from the question of how to recover spectral convergence properties when dealing with conservation laws that exhibit spontaneous shock discontinuities. \citet{Tadmor_SIAM_1989,Tadmor_NASA_1990} studied the use of Fourier-based discretization methods to solve the inviscid Burgers' equations and concluded that, if no regularization term was introduced, the numerical solution would not respect the unique entropy solution and convergence may not be achieved. He then proposed to introduce a wavenumber-dependent viscosity term that would prevent oscillations and lead to convergence to the unique entropy solution. The idea was shown to be successful by mathematical proofs and numerical experiments. Subsequent work by \citet{Karamanos_JCP_2000} and \citet{Pasquetti_JT_2005} applied it to incompressible turbulent flows, via mere addition of the artificial SVV term to the momentum equation without further consideration. Additionally, \citet{Kirby_JFE_2002} applied the same framework to compressible turbulent simulations and artificial spectral dissipation terms were added to the mass, momentum and energy equations. \review{Although the results gathered in these previous works show that the simple extension of the SVV, developed to treat shock discontinuities, to hydrodynamic turbulence works, a clear explanation for the reasons why it worked are lacking.} \citet{Pasquetti_JT_2005} concludes his work on a similar note stating that, although useful, the SVV-LES approach does not rely on physical arguments and therefore it only constitutes an efficient platform with the potential to support an SFS model. The current manuscript addresses this important conceptual gap in section \ref{sec:IncompressibleLES}.

In the current paper, a novel numerical scheme for conducting compressible flow simulations on coarse grids called \modelFullName~(\modelAcronym) is presented. The scheme is based on solving the filtered Navier-Stokes equations using a common mathematical approach to model any type of SFS stresses, whether they are due to turbulence or shocks. \review{In section \ref{sec:IncompressibleLES}, the connection between previous LES and spectral artificial viscosity methods is highlighted. This is done through the analysis of the sub-filter scale terms present in the filtered Burgers' equation and their parallel to those present in the filtered incompressible Navier-Stokes. This establishes the theoretical foundation upon which the \modelFullName\ (\modelAcronym) method is constructed.} Following, section \ref{sec:FourierLagrange} discusses the fully detailed implementation of the \modelAcronym~approach in high-order finite difference solvers focusing on how to estimate the magnitude of the fluctuations near the grid cutoff and on how to introduce a wavenumber modulation to the dissipation spectrum by using the residual of spatial filter operators. Next, in section \ref{sec:LESComp}, the compressible filtered Navier-Stokes equations are presented and the \modelAcronym-based closure models are proposed. \review{Subsequently, section \ref{subsec:ShockCapturing} focuses on demonstrating \modelAcronym's capability of performing simulations of shock-dominated flows by studying the Sod shock tube problem, the Shu-Osher shock-entropy wave interaction, a shock/vortex interaction and shock reflection off a sinusoidal wall. Consecutively, section \ref{sec:QSVturb} discusses \modelAcronym's ability of acting as a turbulence model by analyzing a subsonic Taylor Green Vortex (TGV) test case and compressible turbulent channel flow simulations up to hypersonic bulk Mach numbers. Ultimately, section \ref{sec:QSVboth} assesses the claim of  \modelAcronym~being a unified approach for shock capturing and turbulence modeling by examining the results obtained from a supersonic TGV test case, exhibiting shock-turbulence interaction dynamics.} 



\review{\section{Foundations of a united framework for for SFS turbulence modeling and spectral shock capturing}\label{sec:IncompressibleLES}} 

The objective of this section is to highlight overlooked mathematical similarities between previous eddy-viscosity models and the discontinuity regularization method based on artificial addition of a spectrally vanishing viscosity (SVV)  \citep{Tadmor_SIAM_1989,Tadmor_NASA_1990}. \review{The new perspective presented hereafter serves as a theoretical justification for the unification of the modeling for hydrodynamic turbulence and shock discontinuities, as well as a base for building the \modelAcronym~method, the application of the current idea to high-order finite difference solvers.} \\

\review{\subsection{Similarities between the filtered incompressible Navier-Stokes and the filtered Burgers' equations}}

First, focus is given to hydrodynamic turbulence and its mathematical affinity to wave steepening and discontinuity formation. The normalized incompressible Navier-Stokes are taken under consideration:

\begin{equation} 
\frac{\partial u_i}{\partial x_i} = 0,
\end{equation}

\begin{equation} 
\frac{\partial u_i}{\partial t} + \frac{\partial u_i u_j}{\partial x_j} = - \frac{\partial \mathcal{P}}{\partial x_i} + \frac{1}{Re}  \frac{\partial^2 u_i}{\partial x_j \partial x_j},
\end{equation}

\noindent where $\mathcal{P} = p/\rho_{\text{ref}} U_{\text{ref}}^2$ and $u_i$ are the velocity components nondimensionalized by $U_{\text{ref}}$. These equations are then filtered by an operation that commutes with the derivation,

\begin{equation} \label{eqn:filter}
\overline{f}({\bf x}) = \int f({\bf x}')\overline{G}({\bf x},{\bf x}')d{\bf x}',
\end{equation}

\noindent with an associated filter width ($\overline{\Delta}$) resulting in the filtered incompressible Navier Stokes:

\begin{equation} 
\frac{\partial \overline u_i}{\partial x_i} = 0,
\end{equation}

\begin{equation} \label{eqn:IncompressibleLES}
\frac{\partial \overline u_i}{\partial t} + \frac{\partial \overline u_i \overline u_j}{\partial x_j} = - \frac{\partial \overline {\mathcal{P}}}{\partial x_i} + \frac{1}{Re}  \frac{\partial^2 \overline u_i}{\partial x_j \partial x_j}  - \frac{\partial \tau_{ij}}{\partial x_j},
\end{equation}

\noindent where, $\tau_{ij} = \overline{u_i u_j} - \overline u_i \overline u_j$ is the subfilter scale (SFS) stress tensor, a remainder of the filtering operation applied to the nonlinear governing equations. Since the SFS term depends on the unresolved scales in the flow, it must be modeled. The modeling hypothesis is that the energy flux between the resolved scales and the subfilter scales can be parametrized as akin to a momentum diffusion process:

\begin{equation} 
	- \frac{\partial \tau^d_{ij}}{\partial x_j} = \frac{\partial }{\partial x_j} 2\nu_t\overline S_{ij}, \quad  \tau^d_{ij} = \tau_{ij}  - \frac{1}{3}  \tau_{kk}\delta_{ij },  \quad \overline S_{ij} = \frac{1}{2}\left(\frac{\partial \overline u_i}{\partial x_j} + \frac{\partial \overline u_j}{\partial x_i} \right),
\end{equation}

\noindent  where the superscript $`d'$ indicates the deviatoric components and where $\nu_t$ is the eddy viscosity,  previously discussed in section \ref{sec:IntroLES}. In incompressible LES, the trace of the subfilter stress tensor ($\tau_{kk}$) is absorbed into the pressure term. 
Now, if one uses the same framework to derive the filtered version of the inviscid Burgers' equation, a prototypical representation of nonlinear scalar conservation laws that develops a discontinuity in a finite time, the result is:

\begin{equation}  \label{eqn:FiltBurgers}
\frac{\partial u_1}{\partial t} + \frac{1}{2} \frac{\partial u_1 u_1}{\partial x_1} = 0,  \quad 
 \frac{\partial \overline u_1}{\partial t} + \frac{1}{2} \frac{\partial \overline u_1 \overline u_1}{\partial x_1} =  -  \frac{1}{2} \frac{\partial \tau_{11}}{\partial x_1}.
\end{equation}

\noindent Because of its simplicity, its parallel with the LES framework and its connection to solutions with discontinuities, the Burgers' equation will be used as a test case for the Dynamic Smagorinsky (DYN), the Spectral Eddy Viscosity (SEV) and the SVV models. Previously, an overview of these models and their connection is provided. \\

\subsection{Unified mathematical formulation for eddy and spectral artificial viscosity methods}

One starts by introducing the \citet{GermanoPMC91}'s dynamic procedure (DYN), a way of instantaneously modulating the intensity of the SFS terms by comparing the energy content present in fields filtered with different strengths as an strategy to estimate the energy content of the smallest resolved scales. This extra step leads to the addition of a modulating factor $C$, active only when the scales present in the flow surpass the threshold of the stronger (test) filter. This resolves issues typically associated with the plain Smagorinsky model, which introduces excessive damping during flow transition and does not vanish at the boundaries in wall-bounded flows. In a simplified equation format it can be written as

\begin{equation} \label{eqn:smagorinsky}
\tau^d_{ij} = -  2 C \overline \Delta^2 |\overline S_{ij}|  \overline S_{ij}.
\end{equation}

Another model of interest is the Spectral Eddy Viscosity (SEV) model proposed by \citet{chollet1981parameterization}. First, one starts with the sharp spectral filtered momentum equation in Fourier space,

\begin{equation} \label{eqn:NSFreqDomain}
\left[ \frac{\partial}{\partial t}   + (\nu + \nu_{t}(k,k_c))k^2 \right]  \widehat{ u}(k,t) = t _{< k_c}(k,t),
\end{equation}

\noindent for which the original model was developed. Here,  $t _{< k_c}(k,t)$ is the nonlinear triadic interaction among wavenumbers $k$, $p$ and $q$ such that $k, p, q < k_c$, i.e. nonlinear interaction between resolved scales. Additionally, the energy transfer to the sub-filter scales is modeled via an eddy viscosity term ($\nu_t$) that depends on the wavenumber ($k$) and its cutoff ($k_c$). \citet{chollet1981parameterization} analyzed the energy transfer assuming a Kolmogorov spectrum and modeled it as

\begin{equation}  \label{eqn:spectralEddyViscosity}
 \nu_{t}(k,k_c) = \nu^+_{t}(k/k_c) \sqrt{\frac{E_{k_c}}{k_c}}, \quad \mathrm{where} \quad \nu^+_{t}(k/k_c) =  0.267 + 9.21e^{-3.03 k_c/k},
 \end{equation}
 
\noindent which displays ``plateau-cusp''  behavior as a function of the wavenumber. Moreover, analyzing the dimensionality of the term  $\sqrt{E_{k_c}/k_c}$, one concludes that it comprises the product of a length scale ($\ell = 1/k_c  \approx  2 \overline \Delta$) and a velocity scale ($\upsilon_c =  \sqrt{k_c E_{k_c}}$) related to the motion of the scales near the cutoff. 

If one choses to represent the filtered momentum equation in Fourier space \eqref{eqn:NSFreqDomain} in the physical domain, as in \eqref{eqn:IncompressibleLES}, an equivalent SFS stress tensor 

\begin{equation} 
- \tau^d_{ij}= \sqrt{\frac{E_{k_c}}{k_c}} \left(\mathcal{F}^{-1}[\nu^+_{t}(k/k_c)] * \overline S_{ij} \right),
\end{equation}	 

\noindent is reached where $*$ is the convolution operator and $\mathcal{F}^{-1}$ is the inverse Fourier transform operation. The above relation also benefits from a change in perspective since it can be also rewritten as function of the residual of a low-band pass filter, i.e. a test filter ($\widetilde{G}$), and as function of the local filter width $\overline \Delta$,

\begin{equation} \label{eqn:tauGtilde}
-  \tau^d_{ij} = \sqrt{2 \overline \Delta E_{k_c}} \left( (1 - \widetilde{G}) * \overline S_{ij} \right).
\end{equation}

\noindent  \review{Note that the reformulation of the SEV model in the physical space makes its connection to \citet{GermanoPMC91}'s dynamic procedure more clear: both models effectively apply test filters on the resolved scales to inform the modeling of the unclosed terms. The difference being that the SEV model imposes a \emph{plateau-cusp} shape to the residual of its related test filter by working in spectral space, whereas test filters are used in the dynamic procedure only to control the dissipation magnitude.

Additionally, such a perspective change can elucidate that fact that, if the residual of a filtering operation can be used as a wavenumber modulation function for the eddy viscosity, one would be able to implement such a model as a function of physical space operators rendering it possible to be implemented in high-order finite difference frameworks. This is one of the key elements of the \modelFullName's (\modelAcronym) approach, which will be further discussed in section \ref{sec:FourierLagrange}. }

Ultimately the Spectral Vanishing Viscosity (SVV) \citep{Tadmor_SIAM_1989,Tadmor_NASA_1990} method is addressed. It consists of adding an artificial dissipation term to the Burgers' equation with the aim to regularize its solutions and enable spectral convergence properties away from the discontinuity region. The method reads in physical space as,

\begin{equation} \label{eqn:SVVBurgers}
\frac{\partial  u}{\partial t} + \frac{1}{2} \frac{\partial  u  u}{\partial x} =    \epsilon \frac{\partial}{\partial x} \left(Q * \frac{\partial u}{\partial x}\right),
\end{equation}

\noindent where $Q$ is a viscosity kernel. \review{Looking at such a formulation, one notes its similarity with the filtered Burgers' equation \eqref{eqn:FiltBurgers}, derived using the LES-based idea of solving only for the large scales and modeling the energy flux to the sub-filter scales (SFS). In conclusion, although previously unnoticed, the addition of the SVV artificial viscosity component is akin to the inclusion of an eddy-viscosity term (such as in the previously discussed DYN and SEV models), with the difference being the specific spectral make-up of the modeled unresolved terms.} 

During the development of the SVV model, \citet{Tadmor_SIAM_1989,Tadmor_NASA_1990} applied a Fourier transform to equation \eqref{eqn:SVVBurgers} and proved that the dissipation action by $Q$ leads stability and spectral accuracy away from the discontinuity if

\begin{equation}
 \epsilon \ge \frac{1}{k_c} \quad \mathrm{and} \quad  \mathcal{F}[Q]= \widehat{Q} \geq \mathrm{Const} - \frac{1}{\epsilon k^2}.
\end{equation}

\noindent From these, he interpreted that the minimum ``artificial'' dissipation needed to ensure spectral accuracy would need to be $\epsilon \approx \frac{1}{k_c}$ and could be made to only act in modes above a certain activation wavenumber $m \sim  k_c^{\beta}$, with $\beta < \frac{1}{2}$, rendering most of the spectrum inviscid, as in equation \eqref{eqn:Qhat}. Here it should be pointed out that a formulation where $Q$ affects all the scales, i.e. $m=0$, is also consistent with the theoretical results, as pointed out by \citet{Karamanos_JCP_2000} and that $1/k_c$ is effectively an estimate for the length scale at the cutoff, as pointed out in the previous subsection. 

\citet{Tadmor_SIAM_1989,Tadmor_NASA_1990}'s first proposed SVV implementation was

\begin{equation}  \label{eqn:Qhat}
\widehat{Q} = 
\begin{cases}
     0,& \text{if } |k| \leq  m\\
     1,              & \text{otherwise}
\end{cases}
\end{equation} 

\noindent \review{which we note that, consistently with the spirit of this section, can also be implemented as the residual of a sharp spectral test filter operation.} His results, performed with $m = 2\sqrt{k_c}$, showed that the application of the regularization term made a previously unstable simulation converge. Moreover, it was noted that $C^{\infty}$ smoothness of the viscosity's kernel wavenumber dependency improved the resolution of the method.

In a subsequent work, \citet{Maday_JNA_1993} analyzed the Burgers' equation in the context of a Legendre pseudo-spectral method and proved that the use of a SVV regularization term lead also to convergence to the exact entropy solution. In order to improve SVV's performance, \citet{Maday_JNA_1993} proposed a viscosity kernel of the form

\begin{equation} \label{eqn:SmoothSV}
\widehat{Q}   = 
\begin{cases}
e^{- (k - k_c)^2/(k-m)^2} & \text{if } |k| > m\\
     0,              & \text{otherwise}
\end{cases}
\end{equation}

\noindent where $m = 5 \sqrt{k_c}$ and $\epsilon = k_c^{-1}$. In this form, a vanishing viscosity magnitude, which decreases continuously as the mode number decreases, is obtained. 

In summary, all the presented models, after some changes in perspective, fit into a generalized format of sub-filter scale flux modeling consisting of a magnitude pre-factor composed by a length scale ($\ell$) and velocity a velocity scale ($\upsilon_c$) plus a kernel ($K$) which is convolved with the strain rate tensor ($S_{ij}$), as in 
\begin{equation}\label{eqn:GenTauSGS}
\tau_{ij}= \ell\upsilon_c \left(K * S_{ij} \right).
\end{equation}

\noindent Table \ref{tab:GenLES} emphasizes the connection between the dynamic procedure (DYN), the SEV and the SVV methods and how they relate to the generalized form \eqref{eqn:GenTauSGS}. \review{Table \ref{tab:GenLES} also foreshadows the \modelAcronym's closure (explained in section \ref{sec:FourierLagrange}), which can be loosely interpreted as the extension of the SEV methodology to high-order finite difference implementations in physical space. 
}

\begin{table}[h]
\begin{center}
\begin{tabular}{ c|c|c|c } 
 \hline\hline
  & Length Scale ($\ell$) & Velocity Scale ($\upsilon_c$) & Kernel ($K$)\\ 
 \hline
 DYN &  $\overline \Delta$ &$2 C \overline \Delta |\overline S_{ij}|$ & $1$ \\ 
 SEV & $1/k_c$ & $ \sqrt{k_c E_{k_c}}$ & $\mathcal{F}^{-1}[\nu^+_{t}(k/k_c)]$ \\ 
 SVV & $\epsilon$ & 1& $Q$    \\ \hline

 QSV & $\overline \Delta$ &$\sqrt{2 E_{k_c}/\overline \Delta}$  & $1 - \widetilde{G}_{\text{qsv}}$\\  
 \hline\hline
\end{tabular}
\caption{Summary of different eddy and artificial viscosity models and their connection to a generalized formulation.}
\label{tab:GenLES}
\end{center}
\end{table}

\review{Although the current manuscript is focused on applications based on structured finite difference solvers, the contents of this section may serve as the foundation for the application of an unified approach to shock capturing and SFS turbulence modeling to different platforms, such as unstructured block-spectral solvers.}

\subsection{Analysis of the numerical solution of the filtered Burgers' equation}\label{subsec:Burgers}

The Burgers' equation is integrated in the periodic domain $x_1 \in [-1,1]$, \review{starting from the initial conditions}

\begin{equation} \label{eqn:BurgersInitCond}
u_1(x,t=0) = 1 + \frac{1}{2}\sin(\pi x),
\end{equation}

\noindent until $t = 1$ using a pseudo-spectral Fourier method for the spatial derivatives and a 4th order Runge-Kutta scheme for the time integration. The results obtained using the different eddy viscosity closures are then compared with the analytical solution based on the method of characteristics in the physical and spectral domain. \review{In the current work, the procedure of comparing simulated results to a reference solution will be referred to as {\it a posteriori} analysis. In addition, an {\it a priori} analysis is also conducted. It is defined as an operation that filters the exact solution and its nonlinear terms at a certain time instant by a sharp spectral transfer function down to the same resolution as the numerical simulations performed. This allows the comparison between the exact SFS stress, $\overline{u_i u_j} - \overline u_i \overline u_j$, and the output of each different model when the exact filtered solution $\overline u_i$ is used as input. The intent of this operation is to exploit the access to a reference solution to calculate $\tau_{ij}$ explicitly, the term that needs to be modeled to sustain a filtered solution of a given equation, and assess each model's ability to generate similar effects while only having access to the information present on the large flow scales.}

\begin{figure}[t]
\centering
\includegraphics[width=1.\linewidth]{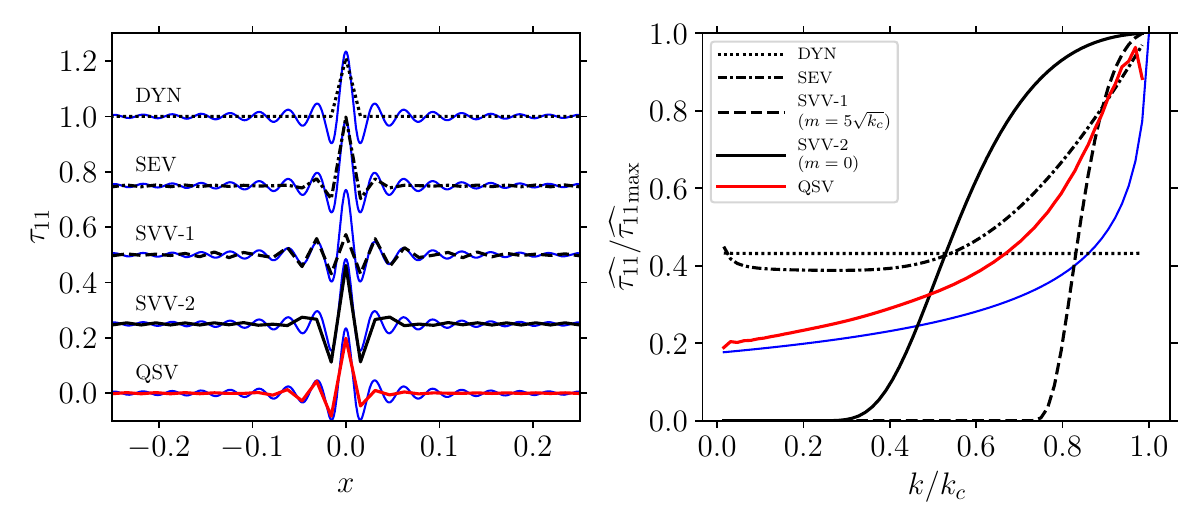}
\put(-260,60){a)} \put(-30,60){b)}\\
\includegraphics[width=1.\linewidth]{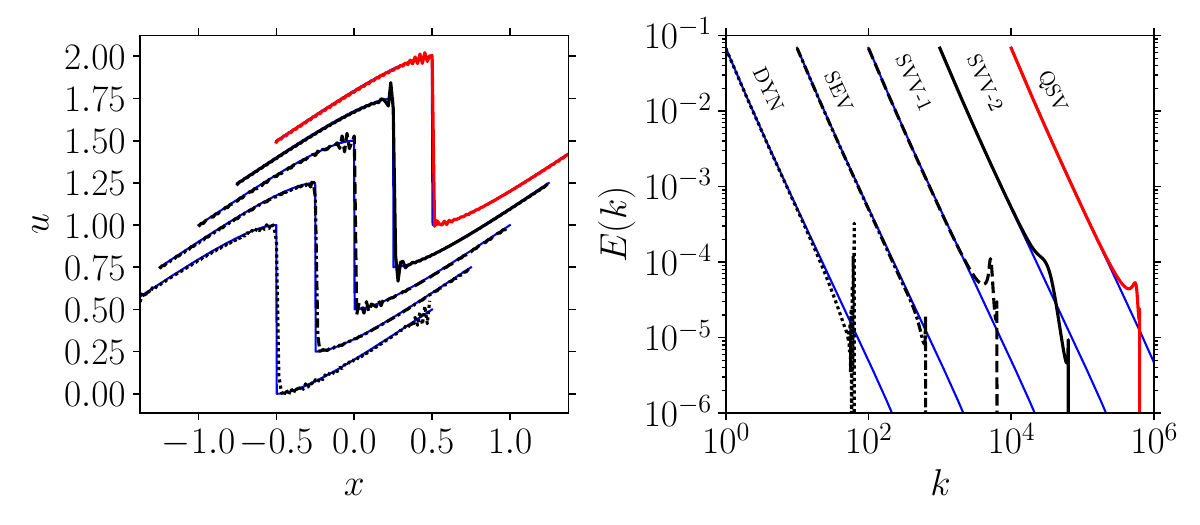}
\put(-400,55){c)} \put(-175,55){d)}
\put(-58,145){
\begin{tikzpicture}
\draw (0,0) node[anchor=north]{}
  -- (0.5,0) node[anchor=west]{\large$-2$}
  -- (0.5,-0.9) node[anchor=north]{}
  -- cycle;
\end{tikzpicture}}
\caption{\review{Results from {\it a priori} and {\it a posteriori} analysis of the filtered Burgers' equation at time $t = 1$ with $N = 128$ grid points. Exact SFS stress (blue) compared against SFS stresses predicted by various models (see legend), in the physical (a) and spectral (b) domain. Exact filtered solution (blue) versus {\it a posteriori} solution by various models (see legend) in the physical space (c) and their respective energy spectrum (d).}}
\label{fig:aprioriPostBurgers} 
\end{figure}
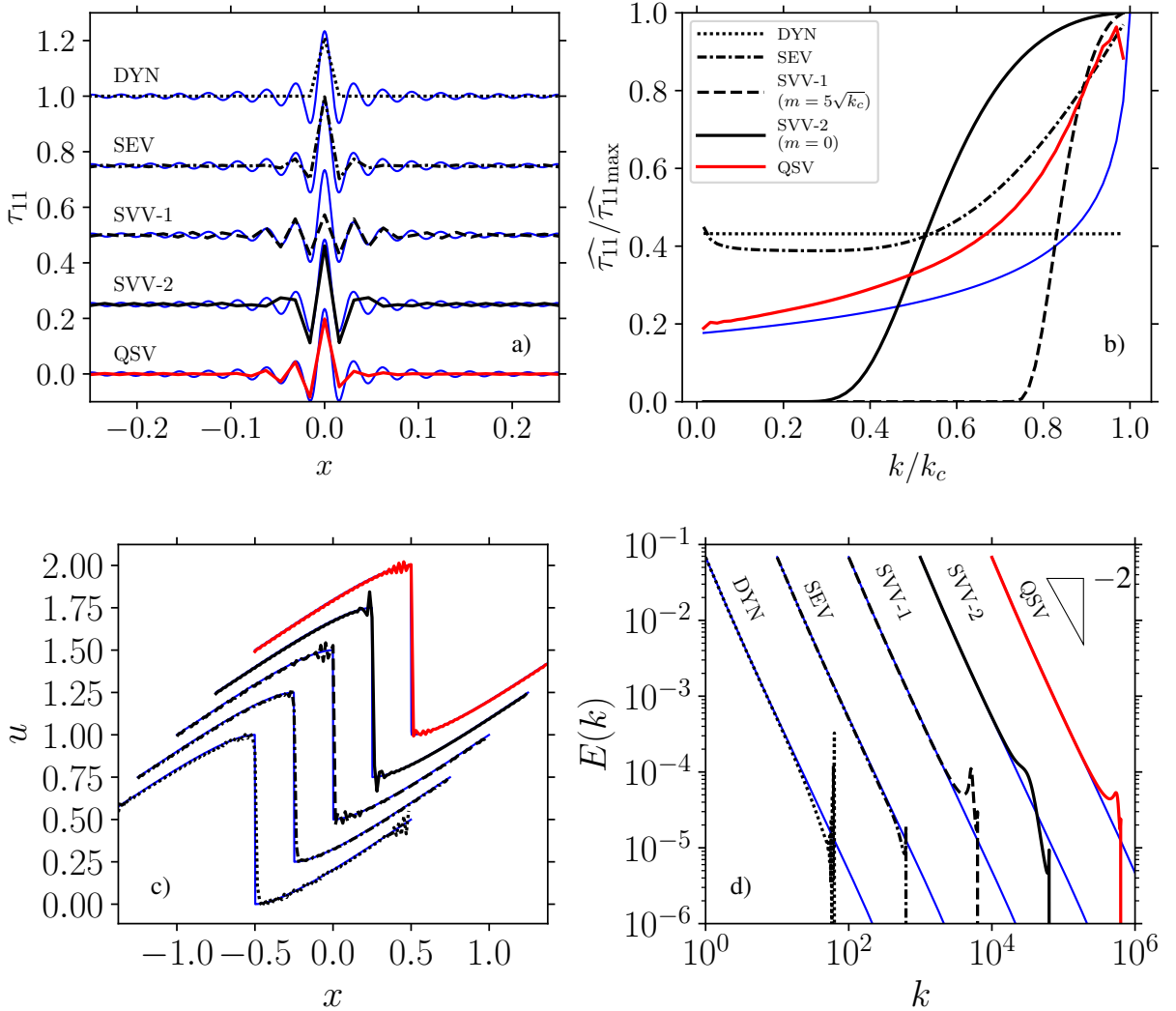


\review{It can be observed, in \ref{fig:aprioriPostBurgers} b) that the exact energy flux to the sub-filter scales (blue line) spans across all resolved wavenumbers and it peaks near the cutoff ($k_c$). This resembles the \emph{plateau-cusp} behavior of the SEV model for hydrodynamic turbulent flows \citep{kraichnan1976eddy,chollet1981parameterization} but observed in the context of a shock-formation phenomena. These results, therefore, can be regarded as another argument in favor of the unification of shock capturing and (hydrodynamic) turbulence modeling methods.}

Now focus is given to the performance of the different models. First, it can be noted in figure \ref{fig:aprioriPostBurgers} that the localization in physical space of the dynamic Smagorinsky procedure leads to a flat broadband response in the wavenumber space that overestimates dissipation at the large scales and underestimates it at scales near the filter's cutoff when modeling the SFS stress. The result of this spectral behavior when used to perform filtered simulations of inviscid Burgers' equation leads to spurious high-wavenumber build up due to the insufficient damping of the resolved scales near the cutoff. This is an undesirable behavior since it affects the overall accuracy of the solution in physical space, as observed by the presence of artificial high frequency oscillations.

Following, the {\it a priori} analysis of the SEV method shows that the growth of the magnitude of the modeled SFS stress components near the grid cutoff is able to follow more closely what is observed in the exact SFS stress although a certain degree of overdamping is introduced. In physical space, the dissipation's ``cusp'' behavior near the cutoff leads to $\tau_{11}$ being slightly non-local, following some of the oscillatory behavior of the analytical result. Moreover, the {\it a posteriori} results show that, in spite of some slight energy build up near the cutoff, the filtered solution follows the energy cascade of the exact solution closely. In the physical space, a mostly monotonic solution is achieved, with only low amplitude high frequency oscillations at the cutoff wavelength.

The SVV method is then analyzed by using both the suggested activation wavenumber  $m = 5 \sqrt{k_c}$ as well as $m = 0$. By studying the model in an {\it a priori} analysis, one can observe that the tentative of leaving part of the spectrum inviscid always leads to underdamping in the low wavenumber range. This reflects in the {\it a posteriori} solution as different degrees of energy accumulation near the cutoff in spectral space and oscillations near the discontinuity in the physical space solution. Moreover, when a larger portion of the spectrum is left inviscid, the level of energy accumulation and amplitude of oscillations are higher. This effect is not limited to the solution of Burgers' equation, as energy pile-up at high wavenumbers was also observed by \citet{Andreassen_JCP_1994}, while using the SVV method to perform simulations of waves in a stratified atmosphere. 

Despite not being the optimal way of introducing a spectral viscosity term, the SVV method does indeed lead to stable and accurate solutions as proved by \citet{Tadmor_SIAM_1989}. SVV paves a strong numerical theoretical background for the implementation of a spectral viscosity term in any hierarchical set of basis functions and its direct connection to the LES mathematical framework clarified in the current manuscript aims to widen its applications and strengthen its technical background when not applied to the initial setup where it was developed. Some examples of implementations of the standard SVV model using different methods are the Fourier \citep{Tadmor_SIAM_1989,Tadmor_NASA_1990}, Legendre \citep{Maday_JNA_1993}, Chebyshev \citet{Andreassen_JCP_1994} and multi domain spectral methods, based on the spectral/hp Galerkin approach, by \citet{Karamanos_JCP_2000}. \review{Similarly, SEV-type models, initially developed in the Fourier spectral space, can be extended to a different spectral orthogonal basis set. This has not been addressed in previous literature.}

\review{One important conclusion that arises from the analysis of figure \ref{fig:aprioriPostBurgers} is that, the closer the spectral content of the added dissipation is to the exact SFS stresses, the better is the performance of the model in solving for the resolved flow scales. Because of that, while developing the \modelAcronym's method, the transfer function of $\widetilde{G}_{\text{qsv}}$ (see table \ref{tab:GenLES}) was tailored to be closer to the exact SFS flux with the added difficulty of only using physical space operators. The mathematical details underlying \modelAcronym's method are discussed in section \ref{sec:FourierLagrange}.  }


Before presenting the inner workings of the \modelAcronym's physical space implementation, it is compared against methods for eddy and artificial viscosity. The analysis of \modelAcronym's results gathered in figure \ref{fig:aprioriPostBurgers} shows a slight energy accumulation at the end of the resolved spectra, near the cutoff wavenumber. Despite the small pile-up, good agreement between the analytical and simulated results are recovered in both physical domain and in the slope of the energy cascade in the low wavenumber range.  When compared against the previous mentioned methods, it performs equally well or better in terms of the range of wavenumbers solved accurately. These results serve as a proof of concept that a spectral-like behavior can be introduced by the use of residuals of filtering operations.



\section{Mathematical formulation of the \modelFullName\ (\modelAcronym) closure} \label{sec:FourierLagrange}

\review{The previous section has established how various eddy and artificial viscosity models can be recast under a common mathematical formulation \eqref{eqn:GenTauSGS}. This demonstrates that the same principles used by large-eddy simulations of hydrodynamic turbulence, focusing on simulating only the large, resolved or filtered scales, can be used for discontinuity capturing as well. Figure \ref{fig:aprioriPostBurgers} from the previous section further establishes this relation and inspires the \modelFullName\ (\modelAcronym) closure, designed specifically to unify the LES and shock capturing methodologies in high-order finite difference implementations. The current section explains the details of its implementation, applied in section \ref{sec:LESComp} to the Favre-filtered Navier-Stokes equations. The following sections demonstrate the model's performance in purely shock-dominated flows, purely turbulent flows and, finally, flows exhibiting shock-turbulence interactions.

The \modelFullName\ (\modelAcronym) closure can be generically written as,

\begin{equation} 
- \tau= \sqrt{2 \overline{\Delta} E_{k_c}} \left( (1 - \widetilde{G}_{\text{qsv}}) * \overline S \right).
\end{equation}	 

\noindent There are two steps for implementing the method: first, estimating the cutoff energy, $E_{k_c}$, and second, introducing a \emph{plateau-cusp} behavior as a function of wavenumber, done by the residual filtering operation, $1 - \widetilde{G}_{\text{qsv}}$. For these to be implemented in a high-order finite difference setting, they need to be performed only using spatial operators, which are discussed in subsections \ref{subsec:Ekc} and \ref{subsec:kModulation}. }

Previously, an attempt to extend the SEV model to physical space was also addressed by \citet{metais1992spectral}, although it only focused on the cutoff energy estimation and not on the spectral modulation. \citet{metais1992spectral} used a structure function to estimate $E_{k_c}$ and averaged the viscosity kernel transfer function to be able to apply a constant coefficient in the wavenumber space. On top of not introducing the ``plateau-cusp'' behavior, the relation between $E_{k_c}$ and the structure function is based on turbulence theory, which assumes local isotropy and homogeneity at the small scales. In the current work, a different approach based on numerical operators, is proposed. Ultimately, the use of numerical theory should relax the assumptions needed to estimate the cutoff energy and also allow the introduction of a wavenumber modulation.

\subsection{Estimation of resolved flow energy at grid cutoff} \label{subsec:Ekc}

The \modelAcronym~method starts by analyzing the residual field associated with a filter based on Pad\'e operators \citep{Lele_JCP_1992}. A family of sixth order implicit Pad\'e filters is defined as 
\begin{equation} 
 \alpha \tilde f_{i-1}+ \tilde f_i + \alpha \tilde f_{i+1} = a_0 f_i + \sum_{j = 1}^{3} \frac{a_j}{2}(f_{i+j} + f_{i-j}), 
\end{equation}

\noindent where $f$ is the quantity being filtered, $\tilde f$ is the filtered field, $\alpha \in (0, 0.5)$ is a parameter which controls the filter's strength  and its weights defined as a function of $\alpha$ are,
\begin{equation}  \label{eqn:PadeWeights}
a_0 = \frac{11 + 10\alpha}{16},\ a_1 = \frac{15 + 34\alpha}{32},\ a_2 = \frac{-3 + 6\alpha}{16}\ \mathrm{and}\ a_3 = \frac{1 - 2\alpha}{32}.
\end{equation} 

\noindent From the filter's definition, one can arrive at its transfer function, 
\begin{equation} \label{eqn:PadeTransferFct}
 \widehat{\widetilde{G}}_{\mathrm{Pade}}\left(\alpha,\frac{k}{k_c}\right) = \frac{a_0 + \sum\limits_{j = 1}^{3} a_j\cos\left(j\pi \frac{k}{k_c}\right)}{1 + 2 \alpha \cos\left(\pi \frac{k}{k_c}\right)},
\end{equation}

\noindent where $k_c = \pi/\Delta$.

\begin{figure}[ht]
\centering
\includegraphics[width=1.\linewidth]{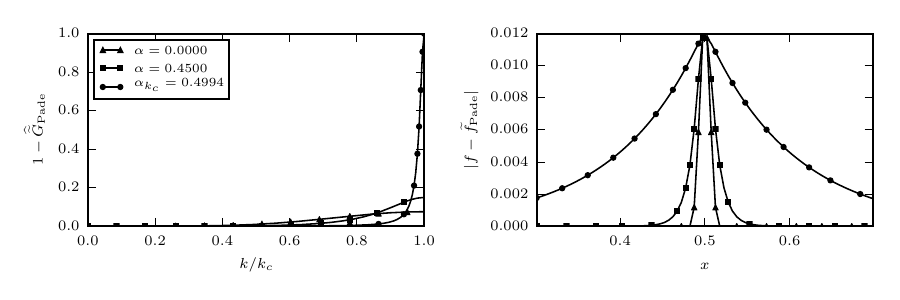}
\caption{Transfer functions obtained for energy at the cutoff operator and the residual of a sixth order Pad\'e filter when $k_c = 64$ (left) are shown together with the physical space result of a Pad\'e residual operation when applied to a unitary step function located at $x=0.5$ (right).
}\label{fig:PadevsFourLagEkc} 
\end{figure}

Figure \ref{fig:PadevsFourLagEkc} shows that the residual of a Pad\'e filter operation can serve as an estimate for the spectral energy content near the cutoff, i.e.
\begin{equation}
\review{ E_{k_c} \approx \left(1 -  \widehat{\widetilde{G}}_{\mathrm{Pade}}\right) * E.}
\end{equation}
 
 \noindent \review{Such residual is a monotonically increasing function of the wavenumber and can be made to concentrate towards the cut-off $k_c$ depending on the value of $\alpha$.} The duality between frequency and physical spaces woven by the Fourier transform, though, leads to the general principle that a function $f(x)$ and its Fourier transform $\hat f(k))$ cannot be simultaneously localized in their own space.  This principle is demonstrated in figure \ref{fig:PadevsFourLagEkc} by showing the physical residual of Pad\'e filters for different values of $\alpha$ applied to a unitary step function located at $x = 0.5$. \review{It can be observed that a higher value of $\alpha$ yields a higher concentration near the cutoff wavenumber in the spectral space, and a corresponding broadening in the physical space, resulting in a (more) global numerical operation. } Ultimately, a trade-off must be found between the accuracy of the estimation of the spectral energy magnitude near the cutoff and the locality of the resulting operation.
 
\review{Initially, only values of $\alpha$ close to $0.5$ result in accurate estimation of the energy content near the cutoff but, as discussed, that leads to highly non-local spatial dissipation. A scaling of the energy at the cutoff is proposed to improve the range of $\alpha$ that leads to good $E_{k_c}$ estimates and, with that, to achieve a higher degree of flexibility in the choice of locality of the energy estimation operation in the physical space.} First, it is noted that the most accurate $E_{k_c}$ estimation for a certain grid discretization using the family of Pad\'e filters is the one for which the inflection point of the transfer function equals to 0.5 at the last resolvable mode prior to the cutoff. This previous statement is equivalent to saying that there exists a maximum $\alpha_{k_c}$, which satisfies the following relation,
 \be
 \widehat{\widetilde{G}}_{\mathrm{Pade}}\left(\alpha_{k_c},\frac{k_c - 1}{k_c}\right) = 0.5.
 \ee

\noindent From here, one can solve for the value of $\alpha_{k_c}$ using relations \eqref{eqn:PadeWeights} and \eqref{eqn:PadeTransferFct}. If  the residual transfer function achieved by using the parameter $\alpha_{k_c}$, $1 - \widehat{\widetilde{G}}_{\mathrm{Pade}}\left(\alpha_{k_c},\frac{k}{k_c}\right)$, is integrated, it is possible to define a reference area connected to the best estimate for the amplitude of the energy near the cutoff for a certain grid size. In general, this integral is only a function of $\alpha$, being 
 \be
A(\alpha) = \int_0^{1}\left(1 - \widehat{\widetilde{G}}_{\mathrm{Pade}}\right) d(k/k_c) =
\begin{cases}
\frac{5}{16} ,& \text{if } \alpha = 0,\\
\frac{ \sqrt{1 - 4\alpha^2} -1 + 2\alpha\left[2 \left(\sqrt{1 - 4\alpha^2} -1 \right) + \alpha\left(14\alpha + 2\sqrt{1 - 4\alpha^2} - 1\right)\right] }{64\alpha^3} ,& \text{if } 0<\alpha < 0.5.\\
\end{cases}
 \ee

Now, if a more localized scheme for energy estimation is desired, one can scale the resulting residual operation by the area ratio between the maximum and desired $\alpha$, reaching the results shown in figure \ref{fig:PadevsFourLagEkc}. \review{Ultimately, in three dimensions, the residual filtering operation can be applied in the $i$-th direction to estimate the cutoff energy content of each resolved $\overline u_j$ velocity component,}

\begin{equation} \label{eq:cutoffEner}
\review{E^i_{k_c} (\overline u_j) = \left[ \left(1 - \widetilde{G^i}_{\mathrm{Pade}}(\alpha)\right)* \frac{(\overline u_j)^2}{2} \right]\frac{A^i(\alpha)}{A^i(\alpha_{k_c})}.}
\end{equation}

\noindent With this formulation, the information on the magnitude is preserved but the use of lower values of $\alpha$ lead to lower wavenumbers, or larger scales, being detected by sensor and again a trade-off must be sought. Although an optimal $\alpha$ might differ for each individual setup, the value of $\alpha =0.45$ produces satisfying results for all the problems in the current manuscript.

\subsection{Wavenumber modulation of the added dissipation} \label{subsec:kModulation}

The question of how to estimate the residual energy in physical space was addressed and now the problem of adding wavenumber dependency on the eddy viscosity term using only spatial operators will be tackled. For this, start by considering the Vandeven filter of order $p=1$ \citep{Vandeven_JSC_1991},
\begin{equation} 
\sigma_1(\eta) = 1 - \eta,
\end{equation}

\noindent defined only on the resolved scales, i.e. $\eta \in [0,1]$. A filter with such a transfer function was studied by \citet{Fejer_1903} in the context of looking for monotonic reconstructions of Fourier series partial sums and will be from now on referred to as Fej\'er filter. Such a filter acts by modulating the amplitudes of the various Fourier components  as

\begin{equation} \label{eqn:SpectralFilter}
f^{\sigma_p} = \sum_{k = -k_c}^{k_c} \hat{f}_k \sigma_p\left(\eta \right) e^{ikx},
\end{equation}

\noindent where $\sigma_p = 1$ results in no filtering. The Fej\'er filter can be cast as a spatial operator. 

Starting from the spectral filter equation \eqref{eqn:SpectralFilter} applied to $u(x)$, one substitutes $\hat u$ for the discrete Fourier transform (DFT) operation,
\begin{equation} \label{eqn:FilterDFT} 
u^{\sigma_1} = \sum_{k = -k_c}^{k_c} \left(\frac{1}{2k_c c_k} \sum_{j = 0}^{2k_c-1} u(x_j)e^{-ikx_j} \right) \left(1 - \frac{k}{k_c}\right) e^{ikx}, 
\end{equation}
\noindent where $c_k$ is a term that appears due to aliasing on the last mode, being equal to $1$ for $|k| < k_c$, and $c_k = 2$ for $k = \pm k_c$ \citep{Shen_SpectralBook_2011}. From this one can derive the operator that, when applied to the discrete values of the function, $u(x_j)$, where $x_j$ are the colocation points $x_j = j \frac{2\pi}{2k_c}$, $0\leq j \leq 2k_c-1$, return its Fej\'er filtered value. From equation \eqref{eqn:FilterDFT} one has that
\begin{equation} 
u^{\sigma_1} = \sum_{j = 0} ^{2k_c-1} \left(\frac{1}{2k_c} \sum_{k = -k_c}^{k_c} \left(1 - \frac{k}{k_c}\right) \frac{1}{c_k} e^{ik(x - x_j)}\right) u(x_j) = \sum_{j = 0} ^{2k_c-1} G_{\text{Fejer},j}( x - x_j) u(x_j).
\end{equation}

 \noindent After some algebraic manipulation and using the help from the closed form of the Fej\'er kernel, the spatial Fej\'er filter operator is defined as 
\begin{equation} 
G_{\text{Fejer},j}(\theta) = \frac{\csc^2\left(\frac{\theta}{2}\right)}{2k_c(2k_c+2)}  \left[ 1 - \cos\left(k_c\theta\right) + \sin\left(k_c\theta\right)\sin(\theta)   \right]
\end{equation}

\noindent where $\theta = x - x_j$. Ultimately, its implementation in a discrete grid where $x_j = j \frac{2 \pi}{2k_c},  j = 0, 1, ..., 2k_c-1$ would be done by solving $u^{\sigma_1} = H_{ij} u_j$ where   
\begin{equation} \label{eqn:spatialFejerFilter}
G_{\text{Fejer},ij} = 
\begin{cases}
    \frac{4 + 2k_c}{2(2+2k_c)},& \text{if } i =  j\\
     \frac{1}{2k_c(2+2k_c)} \csc^2\left(\frac{\pi(i-j)}{2k_c}\right) (1 - (-1)^{i-j}),              & \text{otherwise}.
\end{cases}
\end{equation}

\review{The thus-obtained closed form is strictly speaking a global periodic operator, i.e. weights should be applied to all points in one direction periodically to determine the filtered value at one grid location. This is equivalent to performing a filtering operation in the Fourier spectral space. However, casting it as a spatial operator and truncating its stencil, makes the operation local in space, allowing its extension to non-periodic boundary conditions. }

\begin{figure}[ht]
\centering
\includegraphics[width=1.\linewidth]{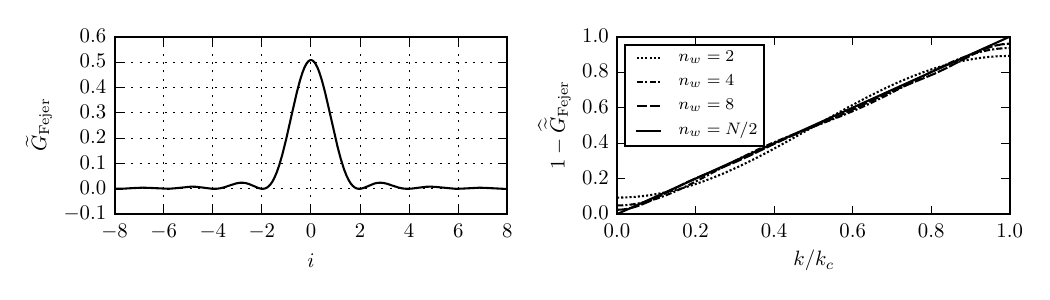}
\caption{ The Fej\'er kernel in space and its self windowed behavior is shown on the left for $N = 128$. On the right, the related transfer function obtained for the global ($n_w = N/2$) and for the truncated operators for different window widths $n_w$.
}\label{fig:TruncFejerFilter} 
\end{figure}

To analyze how the truncation operation affects the Fej\'er filter operator note that, for a centered, explicit and symmetric scheme, generically represented here as,
\begin{equation} 
\widetilde G_i = a_0 G_i + \sum_{j = 1}^{n_w} \frac{a_j}{2}(G_{i+j} + G_{i-j}), 
\end{equation}

\noindent its transfer function can be represented as
\begin{equation}  \label{eqn:GenExpTransferFct}
 \widehat{\widetilde{G}}\left(\frac{k}{k_c}\right) = a_0 + \sum_{j = 1}^{n_w} a_j\cos\left(j\pi \frac{k}{k_c}\right),
\end{equation}

\noindent where $k_c = \pi/\Delta$. Figure \ref{fig:TruncFejerFilter} shows that the self windowed nature of the weights of the Fej\'er filter spatial operator makes its transfer function converge quickly to the theoretical straight line independently of the number of the total number of modes. This behavior renders the truncated version of the operator useful. \review{In the current manuscript a stencil size of $n_w = 8$ is used as the default value, but as the grid points approach a nonperiodic boundary, the window size is decreased accordingly. }

\review{At this point, all the tools necessary to build the effective transfer function used to modulate the magnitude of the spectral dissipation added to each wavenumber are presented. Recall the results presented in figure \ref{fig:aprioriPostBurgers} b): the {\it a priori} SFS dissipation associated with previous models either over or underestimates the exact SFS dissipation at low wavenumbers. On the other hand, the \modelAcronym~spectral modulation transfer function is tailored to be close to the \emph{plateau-cusp} behavior of the exact SFS energy flux. Such modulation is performed by using only spatial operators through a linear combination of the residuals of the truncated spatial Fej\'er filter operator \eqref{eqn:spatialFejerFilter} and the previously mentioned Pad\'e filter \eqref{eqn:PadeTransferFct}: }

\begin{equation}  \label{eqn:EffFilterTF}
\review{\widehat{\widetilde{G}}_{\text{qsv}} = w\left[1 - \beta \widehat{\widetilde{G}}_{\mathrm{Pade}}\left(\alpha\right)  + (1-\beta) \widehat{\widetilde{G}}_{\mathrm{Fejer}}\right].}
\end{equation}

\noindent \review {The values used for the filter strength, $\alpha = 0.45$, for the weighted average factor, $\beta = 0.8$, together with a `DC' offset of $w = 0.85$ were used to generate the results gathered in the current manuscript. 

Moreover, it is necessary to note that how these various parameters interact with each other is ultimately described by the constructed transfer function, shown in figure \ref{fig:aprioriPostBurgers} b), and how each component affects such curve can be translated directly to the model's behavior. Ultimately, changes in the parameters that lead to small changes in the final modulation curve can only generate slight differences in the results achieved by the model.

Although $\alpha,\ \beta$ and $w$ seem to be arbitrarily chosen solely based on the Burgers' test case, their choice has also been informed by other extracting other exact SFS spectra, such as the ones obtained in the {\it a priori} analysis of the Taylor-Green Vortex (figure \ref{fig:TGV_M0-1_apriori}) and the Riemann shock tube problem (figure \ref{fig:aprioriSod}), leading to good agreement and displaying an expected universal behavior, being the \emph{plateau-cusp} shape a theoretical result \citep{kraichnan1976eddy,chollet1981parameterization}. The suggested values for $\alpha,\ \beta$ and $w$ work well for a wide range of test cases, i.e. shock-dominated (section \ref{subsec:ShockCapturing}) or turbulence-dominated (section \ref{sec:QSVturb}) problems, and a shock-turbulence interaction problem (section \ref{sec:QSVboth}), however, they might have to be adjusted if different numerical schemes or filter combinations are adopted. Such values, as well as the choice of spatial filtering operations, may not be optimal, but they shed light on a novel perspective on how to develop eddy-viscosity models. 

Finally, one comment must be made regarding the applicability of the \modelAcronym~closure here derived. Since the spectral modulation is inspired by an {\it a priori} analysis carried out with sharp spectral filters, the model, as constructed, is only suitable for spectral or quasi-spectral high-order discretizations. A different primary filter, for example the transfer function of a second order interpolation operator, could be used in a separate {\it a priori} analysis to look for an alternative QSV implementation in low-order codes.
} 


\section{Application of the \modelFullName\ closure in the filtered compressible Navier-Stokes equations} \label{sec:LESComp}

In this section, the compressible implementation of large scale simulations focusing on the unification of the treatment between shocks and turbulent eddies will be discussed. First the governing equations will be derived from the compressible Navier-Stokes relations using filtering operations and the unclosed terms as well as their closure will be presented. 

\subsection{Governing equations}

The compressible Navier-Stokes system of equations can also be filtered by an operation that commutes with the derivation, as described by \eqref{eqn:filter}, similarly to its incompressible counterpart \eqref{eqn:IncompressibleLES}. After the initial filtering, two routes could be chosen, the Reynolds-based or the Favre-based filtered equations. By using the Reynolds method, all equations in the system would require closure models and the role of small-scale density variations would have to be modeled specifically. \citet{Sidharth_JFM_2018} argued that modeling the SFS flux in the density field separately could lead to an improvement in LES of variable-density turbulence. On the other hand, the use of the Favre-based method leads to implicitly solving density related nonlinear terms by defining the Favre filter operation as,

\begin{equation} 
\check{f} = \frac{\overline{\rho f}}{\overline{\rho}}. 
\end{equation}

\noindent In the end, solving for the Favre filtered quantities $(\check{f})$ simplifies the LES equations for compressible flows. Moreover, compressible LES implementations based on the Favre-filtered equations were already successfully performed with different closure models by, for example, \citet{moin1991dynamic}, \citet{Normand_TCFD_1992}, \citet{vreman1995priori} and \citet{NagarajanLF_JCP_2003}. In this work, we follow the path of the latter, and derive the Favre-filtered Navier-Stokes relations introducing a pressure correction based on the sub-filter contribution to the velocity advection,

\begin{equation} \label{eqn:FavreCont}
\frac{\partial \overline \rho}{\partial t} + \frac{\partial \overline {\rho} \check{u}_j}{\partial x_j} = 0,
\end{equation}

\begin{equation} 
\frac{\partial \overline{\rho} \check{u}_i}{\partial t} + \frac{\partial \overline \rho \check{u}_i \check{u}_j}{\partial x_j} = - \frac{\partial \overline{p}}{\partial x_i} +  \frac{\partial \mu \check{\sigma}_{ij}}{\partial x_j } -  \frac{\partial \overline \rho \tau_{ij}}{\partial x_j},
\end{equation}

\begin{equation} 
\frac{\partial \overline E}{\partial t} + \frac{\partial   (\overline E + \overline p) \check u_j}{\partial x_j} =  \frac{\partial}{\partial x_j} \left(k \frac{\partial \check T}{\partial x_j} \right) +  \frac{\partial \mu \check \sigma_{ij} \check u_i}{\partial x_j }  -  \frac{\partial \overline \rho C_p q_{j} }{\partial x_j}  -  \frac{\partial}{\partial x_j}\left( \frac{1}{2} \overline \rho \mu_{j}\right) +   \mu \epsilon,
\end{equation}

\begin{equation}  \label{eqn:FavreEnergy}
\frac{\overline p} {\gamma - 1} = \overline E -  \frac{1}{2} \overline \rho \check u_i \check u_i - \frac{1}{2} \overline \rho \tau_{ii}.
\end{equation}

The nonlinear terms that contribute to the energy flux from large to small scales are, the SFS stress tensor, 
\be \label{eqn:comptauij}
\tau_{ij} = \widecheck{u_i u_j} - \check u_i \check u_j,
\ee
\noindent the SFS temperature flux,  

\be \label{eqn:compqj}
q_{j} = \widecheck{T u_j} - \check T \check u_j,
\ee
 
\noindent the SFS kinetic energy advection, 
\be \label{eqn:comppi}
\mu_j = \widecheck{u_k u_k u_j} - \check u_k \check u_k\check u_j
\ee

\noindent and the SFS turbulent heat dissipation 
\be \label{eqn:compeps}
\epsilon = \frac{\partial \overline{ \sigma_{ij}  u_i}}{\partial x_j} -  \frac{\partial \check \sigma_{ij} \check u_i}{\partial x_j}.
\ee

\noindent Subfilter contributions resulting from the nonlinearities involving either the molecular viscosity or conductivity's dependency on temperature have been neglected following \citet{vreman1995priori}, who showed those are negligible in comparison against the other terms' magnitudes. Following \citet{NagarajanLF_JCP_2003}, $\mu_j$ and $\epsilon$ are neglected.

\subsection{\modelAcronym's methodology applied to SFS modelling of the compressible Navier-Stokes equations}
\review{
First, the dissipation magnitude tensor is introduced as

\be \label{eqn:Dissipation}
\mathcal{D}_{ij} = \upsilon_i(\check u_j)  \ell_j, 
\ee

\noindent where $\ell_j = \overline \Delta_j$ is the sub-filter length scale and 

\be
\upsilon_i(\check u_j) = \sqrt{ \frac{2 E^i_{k_c}\left(\check u_j \right)}{\overline \Delta_i}}
\ee

\noindent is the sub-filter velocity scale. Note that the cutoff energy estimation operation is carried out in the $i$-th spatial direction (equation \eqref{eq:cutoffEner}) but on the $j$-th filtered velocity component. A complete model for performing compressible large scale simulations that could include both shocks and turbulent events is then proposed as,

\begin{eqnarray} \label{eqn:fullclosure_tau}
\tau_{ij} &=& -  C_{\tau_{ij}} \frac{1}{2} \left( \mathcal{D}_{ij}\frac{\partial \ddot u_i}{\partial x_j} + \mathcal{D}_{ji}\frac{\partial \ddot u_j}{\partial x_i}  \right) ,\\  
 \label{eqn:fullclosure_q}
  q_{j} &=& - C_q  \mathcal{D}_{jj}\frac{\partial \ddot T}{\partial x_j}  , 
\end{eqnarray}

\noindent where the double dot superscript indicates the filter modulated quantities, defined as
\be
 \frac{\partial \ddot u_i}{\partial x_j} =    \frac{\partial \check u_i}{\partial x_j}* \left(1 - \widetilde{G}_{\text{qsv}}\right),
 \ee
 \be
 \frac{\partial \ddot T}{\partial x_j} =  \frac{\partial \check T}{\partial x_j}* \left(1 - \widetilde{G}_{\text{qsv}}\right),
\ee

 \noindent where the modulation step is performed in the three directions and $\widetilde{G}_{\text{qsv}}$ is described in equation \eqref{eqn:EffFilterTF}}. It is possible that the cutoff energy estimation procedure will lead to large variations in space when highly localized flow features, such as shocks, do not align with the grid or the grid itself is deformed. In those cases, a smoothing procedure, i.e. a gaussian filter \citep{Cook_PoF_2007}, can be performed on the $\upsilon_i(\check u_j)$ term.
 
\review{The values of the pre-factors, $C_q$ and $C_{\tau_{ij}}$ are given by:
 \begin{equation}
\quad C_{\tau_{ij}} = 
\begin{cases}
    1.0,& \text{if } i = j,\\
     0.6,& \text{if } i \neq j,
\end{cases} 
\quad \textrm{and}  \quad C_q = 0.8.
\end{equation}
These values are obtained by carrying out a similar {\it a priori} analyses as the one performed to inform the coefficients $\alpha$, $\beta$ and $w$ (see subsection \ref{subsec:Burgers}). The magnitude of these constants are informed, though, through the application of such procedure to test cases such as the 1D Riemann shock tube problem \citep{SOD_1978_JCP} (\ref{sec:AppendixThermalFlux}) and the Taylor Green Vortex (TGV) (section \ref{subsec:TGV}). 



The suggested constants (all of unitary value) work well for a broad set of test cases as shown below. However, if different numerical schemes or filters are used, adjustments might be necessary. In the following, the proposed \modelAcronym's closure will be compared against existing LES and shock capturing strategies in various flow setups ranging from incompressible turbulence to shock-dominated flows, including shock-turbulence interaction test cases.} 



\review{ \section{Demonstrating \modelAcronym's shock capturing capability} \label{subsec:ShockCapturing}} 

\review{Hereafter, \modelAcronym's ability to simulate flows with discontinuities is demonstrated in canonical test cases such as the Riemann shock tube \citep{SOD_1978_JCP}, the Shu-Osher shock-entropy wave \citep{shu1988efficient}, the shock/vortex and shock/sinusoidal-wall interaction problems.} The aforementioned test cases are solved using a 6th-order Pad\'e compact finite difference scheme \citep{Lele_JCP_1992} coupled with a 3rd-order Runge-Kutta time integration method.

\subsection{One-dimensional Shock Dominated Flows} \label{subsec:1DGasDynamics}

\begin{figure}
\centering
\includegraphics[width=1.\linewidth]{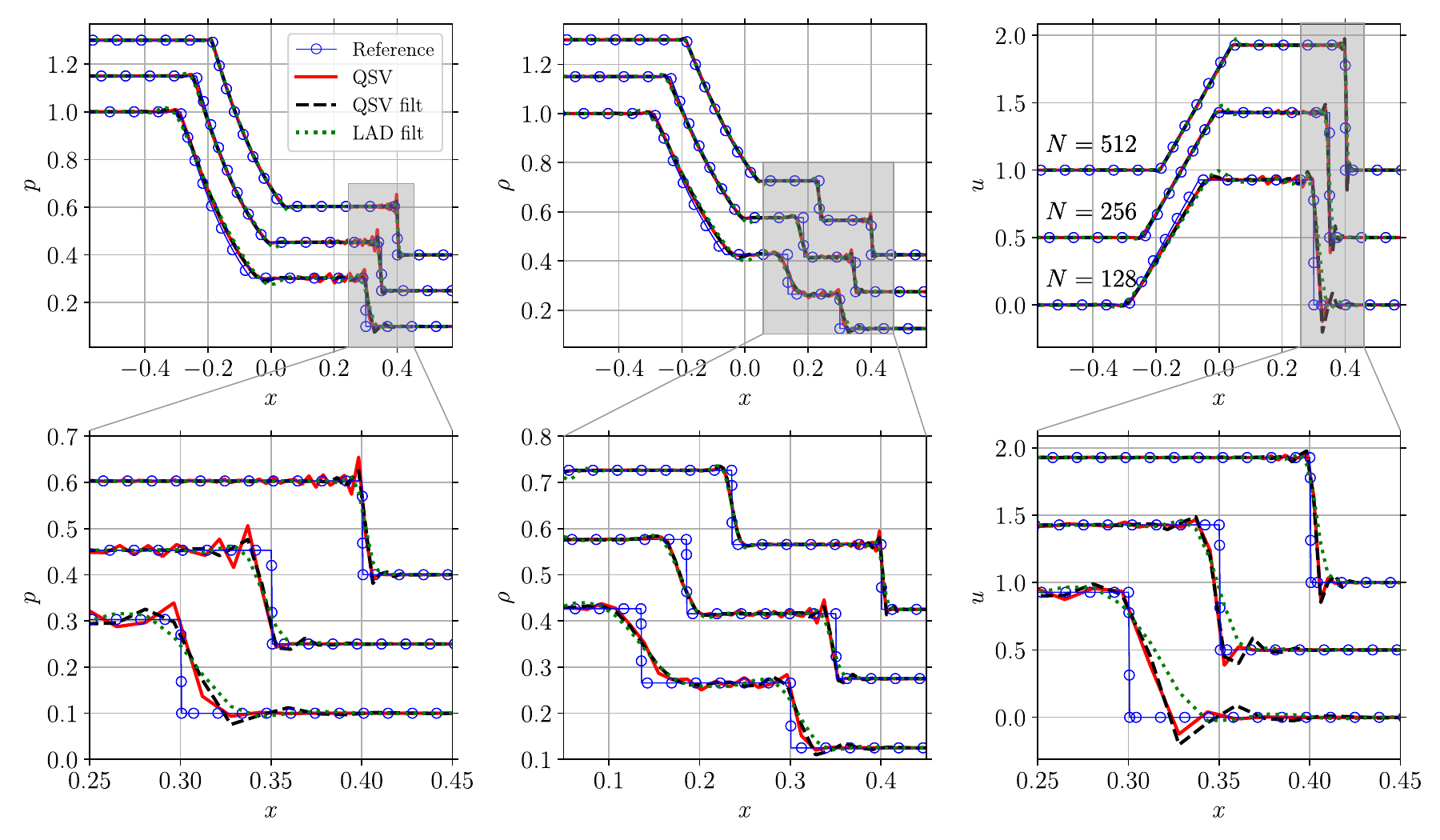}
\caption{\review{Sod shock tube results at $t = 0.2$ computed using the Local Artificial Diffusivity (LAD) method, which requires explicit filtering, and the \modelFullName~(\modelAcronym) method with and without the explicit filtering operation, are compared against the exact solution. }
}\label{fig:SodCFDSU} 
\end{figure}

The Favre-filtered one-dimensional compressible Euler equations read, 

\begin{equation} 
\frac{\partial \overline \rho}{\partial t} + \frac{\partial \overline {\rho} \check{u}_1}{\partial x_1} = 0,
\end{equation}

\begin{equation} 
\frac{\partial \overline{\rho} \check{u}_1}{\partial t} + \frac{\partial \overline{\rho} \check{u}_1 \check{u}_1}{\partial x_1} = - \frac{\partial \overline{p}}{\partial x_1} -  \frac{\partial \overline \rho \tau_{11}}{\partial x_1},
\end{equation}

\begin{equation} 
\frac{\partial \overline E}{\partial t} + \frac{\partial   (\overline E + \overline p) \check u_1}{\partial x_1} = -  \frac{\partial }{\partial x_1} \left(\overline \rho C_p q_{1}\right),
\end{equation}

\begin{equation} 
\frac{\overline p}{\gamma - 1}  = \overline E - \frac{1}{2} \overline \rho \check u_1 \check u_1  - \frac{1}{2} \overline \rho \tau_{11},
\end{equation}

\noindent where the viscous and conductive effects are neglected and where $\tau_{11}$ and $q_1$ are the sub-filter flux terms. If this system of equations is initialized the following initial conditions in density, velocity and pressure,

\begin{equation} 
[\rho,u,p](x,0) = 
\begin{cases}
    [1,0,1],& \text{if } x <  0.0\\
     [0.125,0,0.1],              & \text{otherwise},
\end{cases}
\end{equation}

\noindent then, a single shockwave propagating to the right develops and can be sustained by a Dirichlet supersonic inflow boundary condition on the left and an Neumann outflow boundary condition on the right. As the flow evolves, its exact solution leads to 4 constant density states separated by a shock, a contact discontinuity and a rarefaction wave, as shown in figure \ref{fig:SodCFDSU} at time $t = 0.2$. Shocks and rarefaction regions are also seen in the pressure and velocity fields.

The results for this test case, known as the Sod shock tube problem \citep{SOD_1978_JCP}, obtained via the \modelAcronym~model are then compared against the Local Artificial Diffusivity (LAD) model \citep{Cook_PoF_2007,Kawai_JCP_2008,Kawai_JCP_2010}. Figure \ref{fig:SodCFDSU} gathers a grid sensitivity study of the models considered. \review{Both models appropriately capture the shock and converge to the exact solution as the grid is refined. Nonetheless, the use of the \modelAcronym~model leads to slightly sharper discontinuities for a similar grid refinement level.

The LAD implementation requires an explicit Pad\'e filtering step \citep{Kawai_JCP_2010} with $\alpha = 0.495$, probably due to difficulties with high-wavenumber energy build-up. This step is not required in the \modelAcronym~method, although, if used, a lower degree of filtering (i.e. $\alpha> 0.495$), is already sufficient to control high wavenumber oscillations. The \modelAcronym~model is shown both with and without the application of an explicit filter with $\alpha = 0.499$, in dashed black and solid red, respectively. The explicit filtering step is able to attenuate spurious high wavenumber oscillations that are present in the results. }

Moving forward, another one dimensional canonical problem, the shock-entropy wave interaction \citep{shu1988efficient}, is tackled. The setup comprises a shock wave propagating into a sinusoidally perturbed resting fluid which, upon interaction, are compressed and acoustic waves are generated. The generated waves have sufficient magnitude to induce wave steepening by themselves and ultimately generate a train of weak shocks downstream of the primary shock. This problem is defined by the following initial conditions in the domain $x \in [-1,1]$,

\begin{equation} 
[\rho,u,p](x,0) = 
\begin{cases}
    [3.857143, 2.629369, 10.333333],& \text{if } x <  -0.8\\
     [1.0 + 0.2\sin(5\pi x), 0.0, 1.0],              & \text{otherwise}.
\end{cases}
\end{equation}

\begin{figure}
\centering
\includegraphics[width=1.\linewidth]{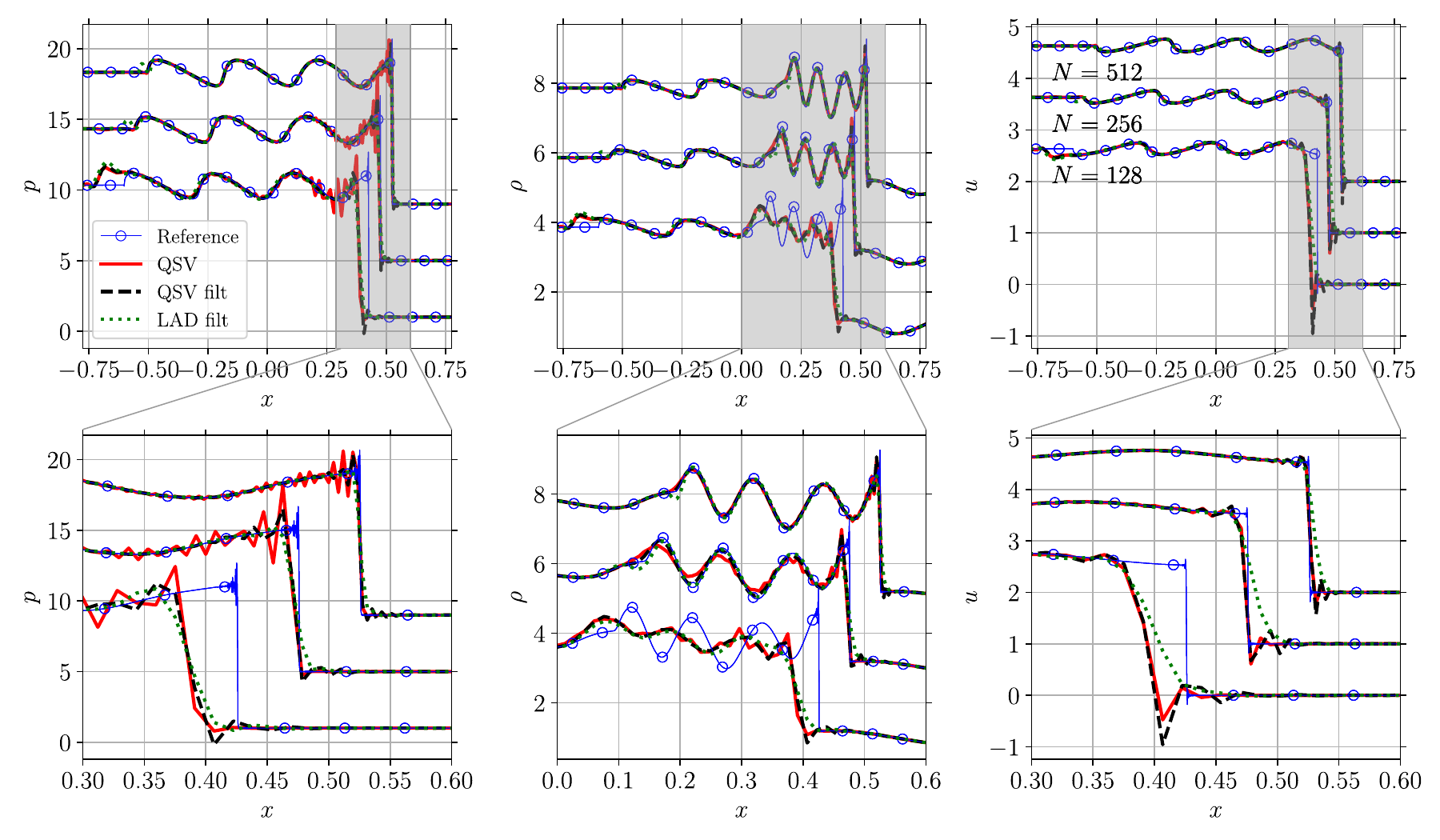}
\caption{\review{Shu-Osher shock entropy wave interaction results at $t = 0.36$ computed using the Local Artificial Diffusivity (LAD) method, which requires explicit filtering, and the \modelFullName~(\modelAcronym) method with and without the explicit filtering operation, are compared against a reference solution computed with 4096 points.}
}\label{fig:ShuOsherCFDSU} 
\end{figure}

\noindent The initial conditions are advanced up until $t = 0.36$ using a timestep of $\delta t = 1\times 10^{-5}$ and results are gathered in figure \ref{fig:ShuOsherCFDSU}.

Once more, results are shown for the \modelAcronym~model with and without the explicit filtering step and for the LAD model using the same color scheme as in figure \ref{fig:SodCFDSU}. \review{Additionally, a \modelAcronym-based simulation using 4096 points is used as a reference solution for the problem. Despite starting from a very coarse grid, a monotonic convergence behavior is observed and the three approaches converge to the reference solution as the grid is refined.  One key aspect to be observed in the simulation of the Shu-Osher problem \citep{shu1988efficient} is how well the trailing-shock density oscillations can be recovered. Both methods display similar performance in that regard. However, use of the \modelAcronym~method leads to a sharper shock discontinuity in comparison with the LAD approach, specially observed in the $u$ profile.}

\subsection{Two dimensional inviscid strong vortex/strong shock interaction}

A stationary shock sustained by an inflow velocity of $V_0 = 1.5\sqrt{\gamma p_0/\rho_0}$ is initialized at $x_s = L/2$ inside a computational domain $\Omega = [0,2L]\times[0,L]$. Superposed to this base state, a compressible zero-circulation vortex is initialized upstream of the shock at $(x_v,y_v) = (L/4,L/2)$ with an inner core radius equal to $a = 0.075L$ and an external radius $b = 0.175L$. This can be translated as ${\bf u } = u_{\theta}(r)  {\bf \hat e}_{\theta} + V_0  {\bf \hat e}_{x}$, where $r$ is the radial distance from the center of the vortex and 

\begin{equation}
 \frac{u_{\theta}(r)}{u_{\theta}(a)} = 
 \begin{cases}
 \frac{r}{a} ,& \text{if } r \leq  a\\
\frac{\eta}{2}\left( \frac{r}{b} -  \frac{b}{r}\right)  ,& \text{if } a < r  \leq b\\
 0,& \text{otherwise,}
 \end{cases}
\end{equation}

\noindent where $\eta = 2(b/a)/[1 - (b/a)^2]$ and the maximum tangential velocity is set to $u_{\theta}(a) = 0.9V_0$. Following the pressure field is initialized so that its gradient balances the centripetal force and the following system of equations is solved based on the ideal gas relation and isentropic compression,

\begin{equation}
\frac{\partial P}{\partial r} = \rho \frac{{u}_{\theta}^2(r)}{r}, \quad P = \rho R T, \quad \frac{P}{P_0} = \left(\frac{\rho}{\rho_0}\right)^{\gamma}.
\end{equation}

Such a setup was previously introduced by \citet{Ellzey_PoF_1995} to study the structure of the acoustic field generated by the shock-vortex interaction, by \citet{Rault_JSC_2003} to analyze the driving mechanisms for the production of vorticity in the interaction at high Mach Numbers and by \citet{Tonicello_CandF_2020} to investigate shock capturing techniques in high-order methods focused on their influence on the entropy field and its non monotonic profile across a shock.

\begin{figure}[ht]
\centering
\includegraphics[width=1.\linewidth]{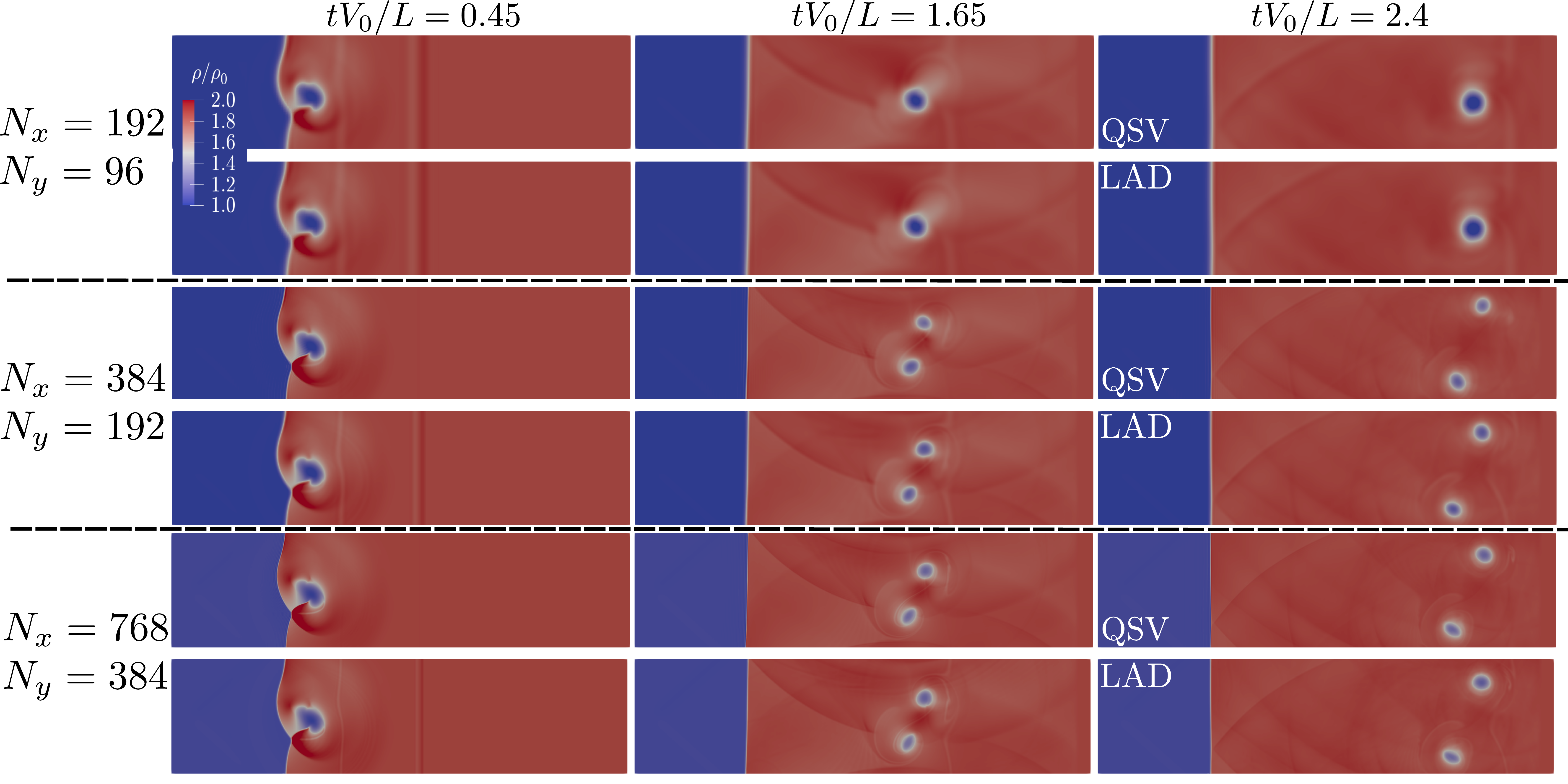}
\caption{\review{Density isocontours of an inviscid shock/vortex interaction for different grid resolutions for both the Local Artificial Diffusivity (LAD) and the \modelFullName~(\modelAcronym) methodology.}
}\label{fig:shockvortex} 
\end{figure}

In the current work, this setup is used to assess the accuracy of the proposed \modelFullName~(\modelAcronym) method. With that objective, the inviscid shock/vortex interaction was solved at 3 different grid resolution levels, $N_x=192;N_y=96$, $N_x=384;N_y=192$ and $N_x=768;N_y=384$, using both \modelAcronym~and the LAD shock capturing scheme \citep{Kawai_JCP_2010}, being the results gathered in figure \ref{fig:shockvortex}. \review{Consistently with \citet{Rault_JSC_2003} and \citet{Tonicello_CandF_2020}, a highly resolved simulation of such flow reveals that such a strong shock/strong vortex interaction is a symmetry breaking event resulting in the formation of two separate  counter-clockwise rotating vortices where the bottom one trails behind the top vortex.

The \modelAcronym~and the LAD models display a similar performance, not being able to capture the asymmetric vortex splitting event on the coarsest grid but yielding a qualitatively correct solution at the intermediate resolutions. Ultimately, both methods converge to the same result, as shown in the bottom row of figure \ref{fig:shockvortex}. Despite the similarities, some differences can be spotted. The \modelAcronym-based results display a sharper shock front and lower noise than the ones performed with LAD for grid sizes $N_x=384;N_y=192$ and $N_x=768;N_y=384$. 
However, the \modelAcronym~model underestimates the post-shock top vortex core size on the intermediate grid with respect to LAD.}

\begin{figure}[ht]
\centering
\includegraphics[width=1.\linewidth]{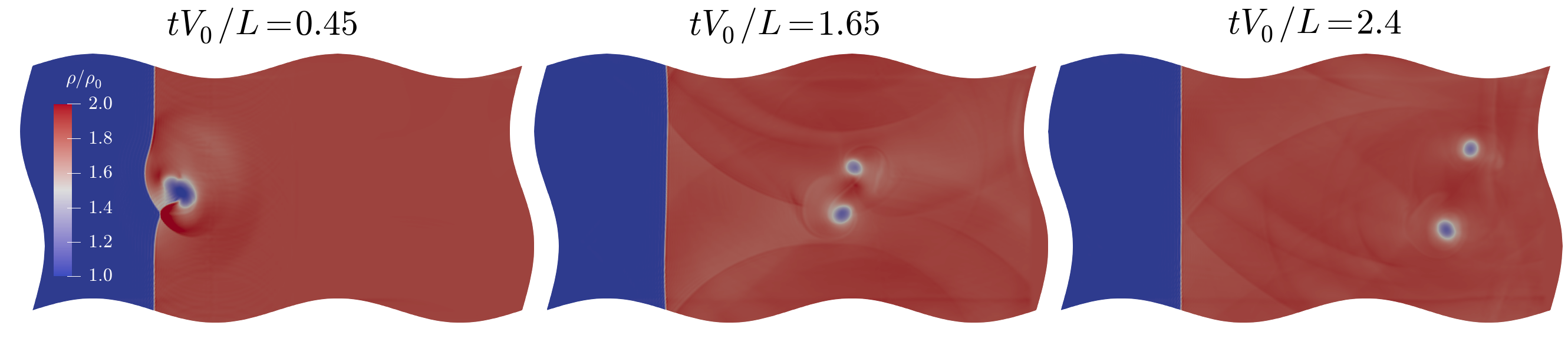}
\caption{\review{Density isocontours of an inviscid shock/vortex interaction using a distorted wavy grid with $N_x = 384$ and $N_y = 192$ points using the \modelFullName~(\modelAcronym) method.}
}\label{fig:wavyshockvortex} 
\end{figure}

Following, the original computational domain for the shock-wave/vortex interaction test case, $\Omega = [0,2L]\times[0,L]$, was distorted by adding a sinusoidal component in both directions where the following linear mapping, 
\begin{equation}
 (x_d,y_d) = \left(x + 0.05L\sin\left(\frac{2\pi}{L}y\right), y + 0.05L\sin\left(\frac{2\pi}{L}x\right)\right)
\end{equation}
\noindent relates the original $(x,y)$ to the distorted grid $(x_d,y_d)$. To solve the compressible Navier-Stokes system of equations with the added closure terms from the \modelAcronym~model in such a nonorthogonal grid, a curvilinear grid transformation such as in \citet{nagarajan2007leading}, was used. The full grid transformed equations and sub filter models are gathered in \ref{sec:AppendixB}.  Figure \ref{fig:wavyshockvortex} shows that \modelAcronym~model is capable of both sustaining the shock and producing qualitatively accurate results even in distorted grids.

\subsection{Two dimensional sinusoidal wall/shock-wave interaction}

The reflection of a shock wave from a sinusoidal wall was chosen as a test case to assess how the \modelAcronym~model behaves in the presence of solid boundaries and compare it against the LAD approach. The numerical setup is built with the objective to mimic experiments reported by \citet{Denet_CST_2015}, where a planar shock wave impinges on a sinusoidal wall with $1.0$ mm amplitude and wavelength of $2.0$ cm. \review{These experiments were designed to confirm the theoretical results by \citet{Clavin_JFM_2013} that predicted the formation of a lasting pattern of triple points after a shock reflection off a smooth sinusoidally-perturbed wall.}

Following \citet{Tonicello_CandF_2020}, a computational domain encompassing one wavelength and $10$ cm in the wall normal direction is used for two numerical experiments. The first is a $M=1.5$ shock propagating in air ($\gamma = 1.4$), similar to the original experiments, and the second is a $M=5.0$ shock propagating in a fluid with a lower specific heat ratio of $\gamma = 1.15$, which increases the strength of the reflected shock and approaches the Newtonian limit \citep{lodato:17}. The simulations are solved with periodic conditions at the top and bottom boundaries together with a no slip adiabatic wall on the left. Moreover, this test case confirms the capability of the \modelAcronym~model to solve the filtered equations in distorted grids.

The shock is initialized $7.5$ cm away from the wall with left and right states corresponding to

\begin{minipage}[l]{.32\textwidth}
\begin{equation} \nonumber
\rho = 
\begin{cases}
\rho_l = 1.208 \text{ kg m}^{-3}, \\
\rho_r = \rho_l \frac{(\gamma + 1)M^2}{2 + (\gamma - 1)M^2},
\end{cases}
\end{equation}
\end{minipage}
\begin{minipage}{.32\textwidth}
\begin{equation} \nonumber
u = 
\begin{cases}
u_l = 0.0 \text{ m s}^{-1},\\
u_r = a_r \sqrt{\frac{(\gamma - 1) M^2 + 2}{2 \gamma M^2 - (\gamma - 1)}} - a_l M ,
\end{cases}
\end{equation}
\end{minipage}
\begin{minipage}[r]{.32\textwidth}
\begin{equation}
p = 
\begin{cases}
p_l = 101.325 \text{ kPa},\\
p_r = p_l \frac{2 \gamma M^2 - (\gamma - 1)}{\gamma + 1},
\end{cases}
\end{equation}
\end{minipage}

\noindent where $a_{l,r}  = \sqrt{\gamma \frac{p_{l,r}}{\rho_{l,r}}}$ is the speed of sound in either the left or right states. In both simulations the Sutherland's law for air was used to model dynamic viscosity.

\begin{figure}[ht]
\centering
\includegraphics[width=1.\linewidth]{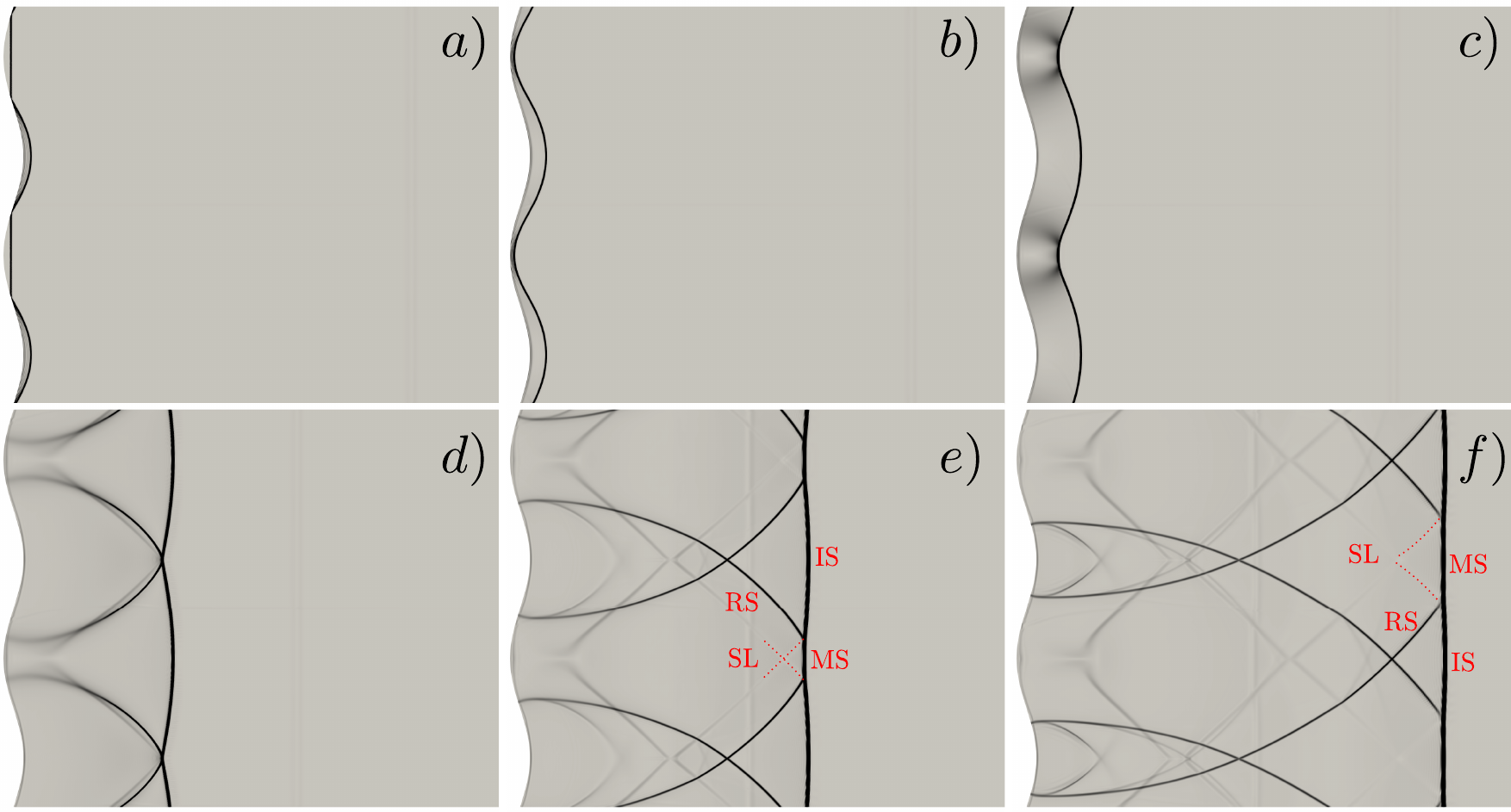}
\caption{\review{Numerical Schlieren visualization of the patterns created by a $M=1.5$ shock reflection off a sinusoidal wall achieved by performing a \modelAcronym~simulation using $N_x = 1024$ and $N_y = 256$ grid points. The shown evolution time frames are: a) $0.0\ \mu$s, b) $2.8\ \mu$s, c) $12.8\ \mu$s, d) $48.5\ \mu$s, e) $88.9\ \mu$s, f) $129.3\ \mu$s. Highlighted in the figure are the incident shock (IS), the reflected shock (RS), the Mach stem (MS) and the slip line (SL). }
}\label{fig:wavywall1-5} 
\end{figure}

Figure \ref{fig:wavywall1-5} gathers the results of the simulation of the reflection of the $M=1.5$ shock reflection when solved with the addition of the \modelAcronym~in a mesh with 1024 points in the wall normal direction and 256 points in the tangential direction. Although a hyperbolic stretched grid was used to concentrate the resolution near the wavy wall, the current resolution is 8 times less fine in terms of degrees of freedom when compared with the simulations presented in \citet{Tonicello_CandF_2020}, for this case. \review{Despite the lower resolution, the results qualitatively agree with previous experimental \citep{Denet_CST_2015} and numerical \citep{lodato:16,lodato:17,Tonicello_CandF_2020} results.

As discussed in \citet{lodato:16}, the small amplitude wall waviness leads to a pressure increase at the valleys and the shock reflection is initially regular. Moments later, triple points, composed by the superposition of the boundaries of four different lines: the incident shock (IS), a reflected shock (RS), the Mach stem (MS) and the slip line (SL), which separates regions of different tangential velocity, are formed. Such points are responsible for the formation of a pattern similar to ones observed in cellular detonations. Moreover, the arrangement of the triple point configuration gets reversed after each collision with another triple point, i.e. the incident shock and mach stem portion are transposed. All of these dynamics are captured by the \modelAcronym~simulation and are represented in figure \ref{fig:wavywall1-5}.
}

\begin{figure}[ht]
\centering
\includegraphics[width=1.\linewidth]{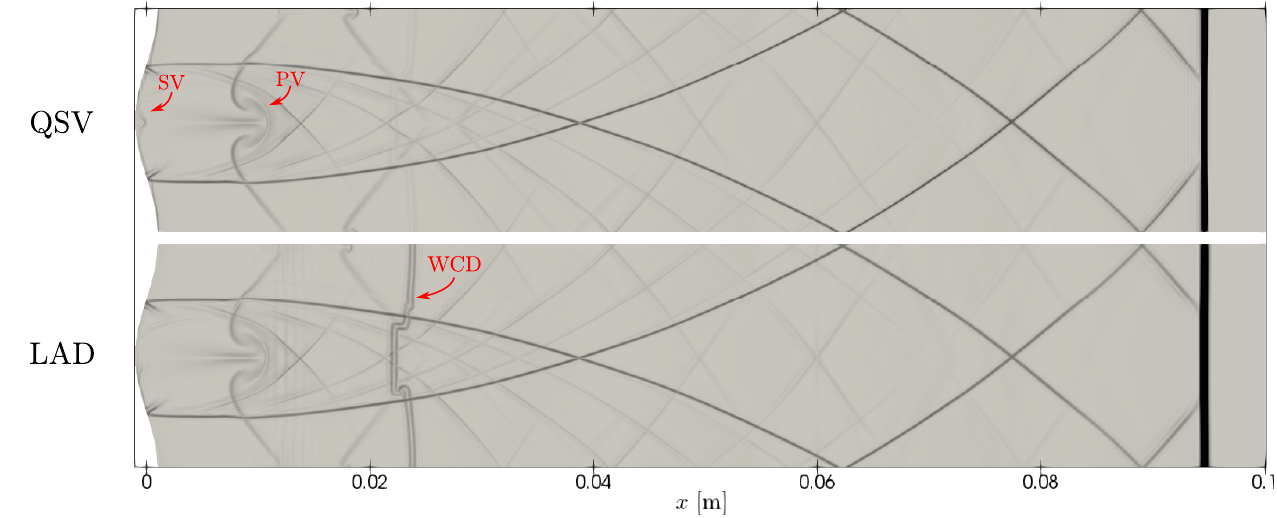}
\caption{\review{Numerical Schlieren visualization \modelAcronym- (top) and LAD-based (bottom) simulations of a $M=1.5$ shock reflection off a sinusoidal wall at $t = 291.5\ \mu$s after the initial shock collision.  The simulations were performed with $N_x = 1024$ and $N_y = 256$ grid points. Highlighted in the figure are the weak contact discontinuity (WCD), the primary vortex (PV) and the secondary vortex (SV).}
}\label{fig:wavywall1-5vsLAD} 
\end{figure}

\review{Figure \ref{fig:wavywall1-5vsLAD} shows the direct comparison between \modelAcronym~ and LAD results for the same test case. First, it can be noticed that a spurious vertical line, a weak contact discontinuity (WCD), is present in both simulations, although being much less pronounced in the \modelAcronym-based one. This artifact is the result of initializing the simulations with a sharp jump between the left and right shock states, and it is also observable in the DNS runs conducted by \citet{lodato:16}. In the first few iterations, the numerical scheme reacts to the ideal sharp discontinuity by diffusing it. This leads to the generation of a right-running weak acoustic wave, which leaves the domain, and a weak contact discontinuity advecting towards the left, which remains in the domain. Ultimately, its strength is not enough to perturb the dynamics of the simulations.

Besides displaying higher intensity of such artificial contact discontinuity, the LAD-based simulation also displays spurious oscillations near the top of the primary vortex (PV), not present in the \modelAcronym~simulation. The near wall region also exhibits some differences. A secondary vortex (SV), also observed in DNS results by \citet{lodato:16} is captured by the \modelAcronym~simulation but is smeared out by the LAD method. Apart from these differences, all the other larger-scale structures are resolved equally well by both methods.}

\begin{figure}[ht]
\centering
\includegraphics[width=1.\linewidth]{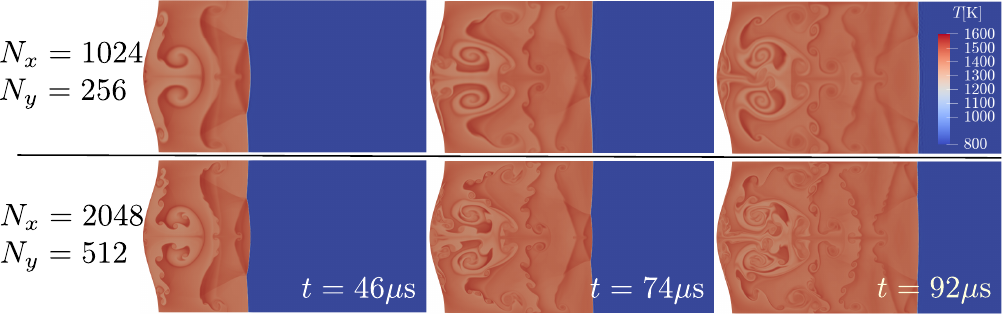}
\caption{\review{Temperature isocontours visualization of the patterns created by a $M=5.0$ shock in a fluid with $\gamma = 1.15$ after it is reflected off a sinusoidal wall at different grid resolutions performed via a \modelAcronym~simulation.}
}\label{fig:wavywall5-0} 
\end{figure}

\review{
Figure \ref{fig:wavywall5-0} shows a similar test case, where the Mach number of the incident shock is increased to $M = 5.0$ and the fluid's specific heat ratio is decreased to $\gamma = 1.15$, so that the test case approaches the Newtonian limit. This case displays stronger variations across discontinuities and higher levels of vorticity along slip lines, with more complex flow patterns generated after shock reflection.  In fact, \citet{lodato:17} reported a twentyfold increase in vorticity magnitude when the Mach number is increased from $1.5$ to $5.0$ and $\gamma$ is decreased from $1.4$ to $1.15$. Moreover, it is possible to observe the presence of triple points at the intersection of the incident shock (IS), the Mach stem (MS), the reflected shock and the slip line (SL), easier to distinguish in figure \ref{fig:wavywall5-0} than in figure \ref{fig:wavywall1-5}  due to their strength. Figure \ref{fig:wavywall5-0} also makes it possible to observe that slip lines are detached post collision and newly ones are formed. After detachment, the tip of the slip lines form strong counter-rotating vortex pairs, similar to those observed behind a detonation front \citep{Bourlioux_CaF_1992}, as pointed out by \citet{lodato:17}. Finally, more flow features are captured, as expected, as the grid is refined.
}

\begin{figure}[ht]
\centering
\includegraphics[width=1.\linewidth]{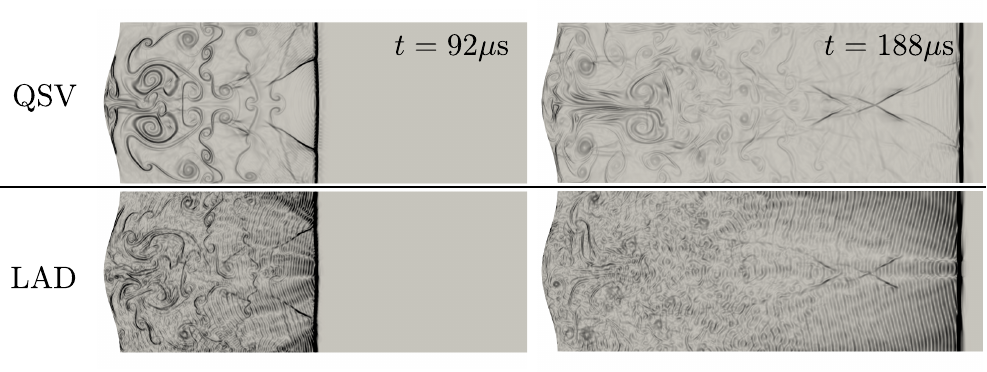}
\caption{\review{Numerical Schlieren comparison between \modelAcronym- and LAD-based simulations of a $M=5.0$ shock in a fluid with $\gamma = 1.15$ after it is reflected off a sinusoidal wall performed with $N_x = 1024$ and $N_y = 256$ grid points}.
}\label{fig:wavywall5-0vsLAD} 
\end{figure}

\review{Figure \ref{fig:wavywall5-0vsLAD} shows a comparison between the resolution capability of \modelAcronym- and LAD-based simulations for the same test case. The \modelAcronym~approach is able to accurately simulate the flow dynamics and preserve symmetry. On the other hand, when the LAD approach is used to solve the current test case, a high level of spurious oscillations was observed. Additionally, an early symmetry breaking behavior is present. Similar post shock oscillations were reported to be a numerical challenge during in the analysis performed in \citet{lodato:17} and inspired the development of an improved shock sensor technique based on using characteristic variables \citep{lodato:19}.
}
\review{
\subsection{Summary}
}
\review{In summary, after the executing one- and two-dimensional test cases related to shock-dominated flows, one can conclude that the \modelAcronym~closure is able to perform shock capturing. Furthermore, the \modelAcronym~model is shown to be able to conduct the aforementioned task accurately, with similar resolution capability as the established LAD model but with higher degrees of robustness in the presence of stronger shock intensities.}
\\

\review{\section{Demonstrating \modelAcronym's sub-filter-scale (SFS) turbulence modeling capability} \label{sec:QSVturb}}

This section focuses on showcasing \modelAcronym's capability of acting as a turbulence model by solving a subsonic compressible Taylor-Green vortex (TGV) setup and a turbulent channel flow up to hypersonic turbulent bulk Mach numbers covering both homogeneous isotropic and wall-bounded turbulence setups. Again, the test cases are solved using a 6th-order Pad\'e compact finite difference scheme \citep{Lele_JCP_1992} coupled with a 3rd-order Runge-Kutta time integration method.
\\
 \subsection{Subsonic Taylor Green Vortex (TGV)} \label{subsec:TGV}

The complete \modelAcronym~closure relations \eqref{eqn:fullclosure_tau} - \eqref{eqn:fullclosure_q} for the filtered compressible Navier-Stokes are now tested by assessing their capability of modeling the energy flux from large to small scales in three-dimensional turbulence problems. \review{Initially, the evolution of a Taylor-Green vortex (TGV) is studied via {\it a priori} and {\it a posteriori} analyses. The latter consists in comparing the result of a coarse simulation started from the initial conditions against a reference solution, in this case a Direct Numerical Simulation (DNS). The former, on the other hand, is based on using the same reference solution and a sharp spectral filter operation to directly obtain the exact value of the unclosed SFS stress term, $\widecheck{u_i u_j} - \check u_i \check u_j$, as well as to evaluate models for the SFS stress field by using the Favre-filtered velocity, $\check u_i$, as the model's input. The exact and modeled SFS stress terms are then compared.}
 
 The TGV test case is defined at $t=0$ in a cubic domain $\Omega \in [-\pi L,\pi L]^3$ as, 

\begin{eqnarray} 
\rho(\mathbf{x},0)/\rho_0 &=& 1,\\
u_1(\mathbf{x},0)/V_0 &=&  \sin\left(\frac{x_1}{L}\right) \cos\left(\frac{x_2}{L}\right) \cos\left(\frac{x_3}{L}\right),\\
u_2(\mathbf{x},0)/V_0 &=& - \cos\left(\frac{x_1}{L}\right) \sin\left(\frac{x_2}{L}\right) \cos\left(\frac{x_3}{L}\right),\\
u_3(\mathbf{x},0)/V_0 &=& 0,\\
p(\mathbf{x},0)/\rho_0 V_0^2 &=&  \frac{p_0}{\rho_0 V_0^2} + \frac{1}{16} \left[\cos\left(\frac{2 x_1}{L}\right)+\cos\left(\frac{2 x_2}{L}\right)\right] 
\left(\cos\left(\frac{2 x_3}{L}\right) + 2\right),
\end{eqnarray}

\noindent where $\rho_0,V_0$ and $L$ are used as nondimensionalization constants. For these simulations $\rho_0$ and $p_0$ are set to $1$ and $1/\gamma$, respectively. Additionally, $V_0$ is used as a parameter that controls the initial fluctuation Mach number $M_t = V_0/\sqrt{\gamma p_0/\rho_0} $ and, therefore, the compressibility effects. In the current manuscript, the Reynolds number $Re = \rho_0 V_0 L /\mu_0$ is set to 5000, independently of the chosen Mach number. 

\begin{figure}[ht]
\centering
\includegraphics[width=1.\linewidth]{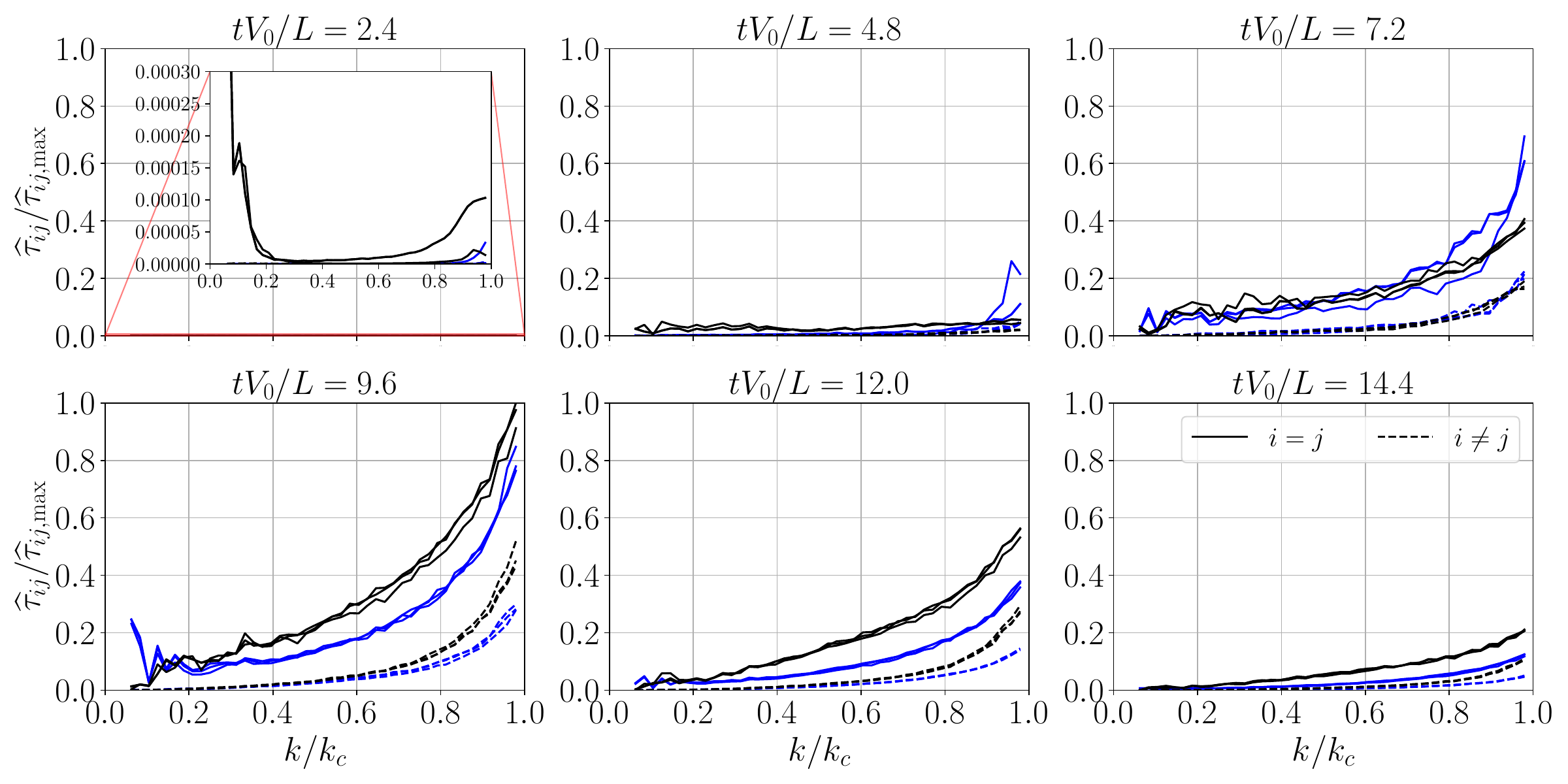}
\caption{\review{{\it A priori} analysis conducted on a $M_t = 0.1$ TGV reference solution obtained via Direct Numerical Simulation (DNS). A sharp spectral filter with cutoff $k_c = 48$ is applied both to evaluate the exact SFS stress tensor $\widecheck{u_i u_j} - \check u_i \check u_j$, shown in blue, and to evaluate \modelAcronym~closure for the modeled SFS stress tensor with the exact filtered velocity field as the input (black lines). Results are compared in the wavenumber space for different times.}
}\label{fig:TGV_M0-1_apriori} 
\end{figure}

\review{Initially, a low Mach number $M_t = 0.1$ is chosen to avoid compressibility effects with the objective to focus only on \modelAcronym's performance when applied to hydrodynamic turbulence. First, an {\it a priori} study is performed and data from a DNS simulated with $384^3$ grid points is used to directly evaluate the spectral content of the exact sub-filter stress tensor $\tau_{ij}$. These results are then compared, in figure \ref{fig:TGV_M0-1_apriori}, against the outcome of the \modelAcronym~model if applied to sharp-spectrally filtered DNS data with grid cutoff $k_c = 48$. }

\review{Figure \ref{fig:TGV_M0-1_apriori} shows the evolution of the different components of the exact and modeled SFS stress tensor at various instants as the initial vortex breaks down and, by doing so, it displays the existence of a distinct behavior between the trace and off-diagonal components. The components that belong to the trace of the exact tensor have a higher magnitude for all wavenumbers up to $k_c$ when compared to the off-diagonal components $\tau_{12},\tau_{13},\tau_{23}$. This justifies the choice to decrease the magnitude in the off-diagonal components, as discussed in the end of section \ref{sec:LESComp}, resulting in the coefficient $C_{\tau_{ij}}$ is set to 0.6 and 1.0 for the off-diagonal and diagonal components, respectively.}

\begin{figure}[ht]
\centering
\includegraphics[width=1.\linewidth]{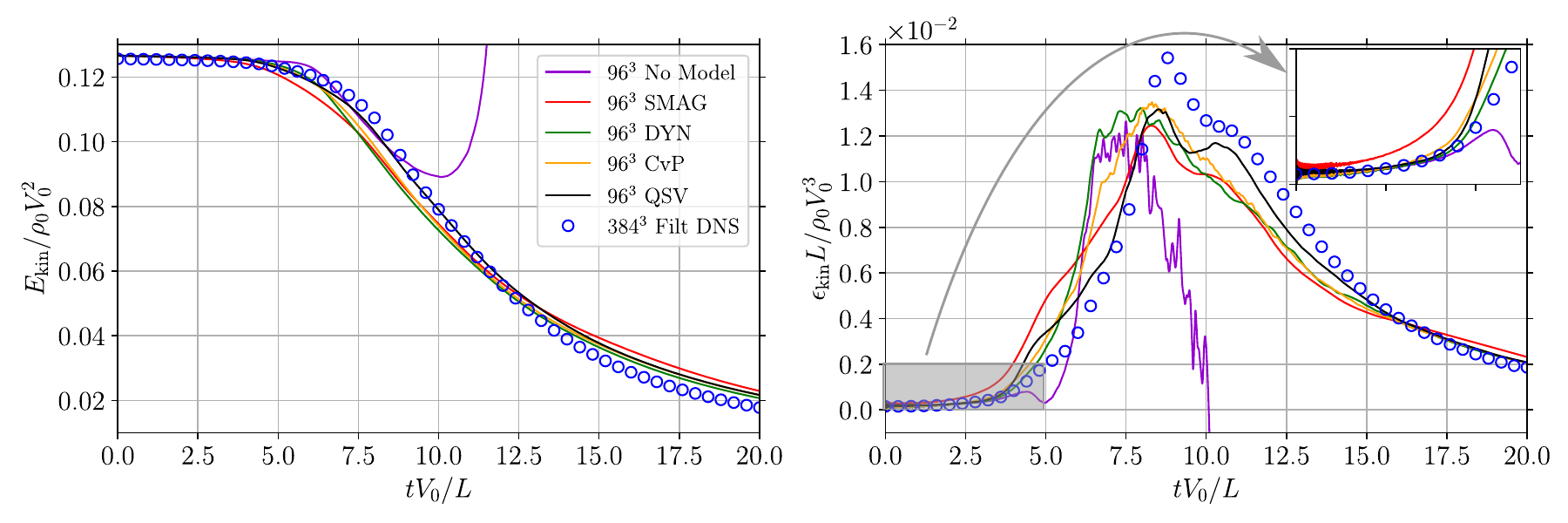}
\put(-430,125){a)}\put(-195,125){b)}
\caption{\review{Evolution of volume-averaged \review{kinetic} energy (a) and dissipation (b) of the LES of the Taylor-Green Vortex with initial amplitude of $M_t = 0.1$. Different subfilter models are shown and compared against sharp spectral filtered DNS data up to the same grid resolution. }
}\label{fig:TGV_M0-1_TKEeps} 
\end{figure}

\review{Furthermore, figure \ref{fig:TGV_M0-1_apriori} displays the dissipation's transient behavior as the initially large vortices breakdown into small-scale turbulence. At the beginning of the simulation, only large scales exist and the dissipation added by the \modelAcronym~model driven by the filtered DNS results, as well as its exact counterpart, are insignificant. As the flow develops, the exact energy flux to sub-filter scales first increases, then reaches a peak and ultimately decreases, as also observed in figure \ref{fig:TGV_M0-1_TKEeps}. At the same time, the operation responsible for estimating energy near the cutoff is able to regulate correctly the magnitude of the SFS dissipation to account for this dynamic behavior. Moreover, the {\it plateau-cusp} behavior is accurately reproduced by the model, which follows closely the directly calculated values.}

\review{An {\it a posteriori} study is then performed with $96^3$ grid points, comparing the \modelAcronym~model, the Smagorinsky model (SMAG), the Dynamic procedure (DYN) and the Coherent vorticity Preserving (CvP) method against the DNS sharp-spectrally filtered down to the LES grid resolution.  Details on the implementation of the latter methods, such as constants and test filter strength, can be found in \citet{Chapelier_JCP_2018}.} Specifically, the state of the turbulence of the different TGV simulations is monitored by analyzing the evolution of volume-averaged kinetic energy, \review{$E_{\mathrm{kin}} = \langle\frac{1}{2} { \rho u^2}\rangle$}, and its dissipation rate, defined as,

\begin{equation}
\epsilon_{\mathrm{kin}} = - \frac{\partial E_{\mathrm{kin}}}{\partial t},
\end{equation}

\noindent with results gathered in figure \ref{fig:TGV_M0-1_TKEeps}. 

\review{Simulating a $M_t = 0.1$ TGV flow with $96^3$ points without any turbulence model leads to a numerically unstable run. In the absence of a model, $\epsilon_{\mathrm{kin}}$ becomes positive around $tV_0/L \approx 10$, indicating a spurious generation of kinetic energy, ultimately leading to diverging numerical results. Nonetheless, the results without the addition of extra dissipation to account for the energy flux to sub-filter scales can be used to assess the performance of SFS models in the early stages of the simulation, when only large scales exist and the models should be inactive.

In figure \ref{fig:TGV_M0-1_TKEeps}, a zoomed region focused on the period $tV_0/L = 0 - 5$ shows that all the models considered, apart from the plain Smagorinsky model, are able to mitigate the addition of excess dissipation in the early stages of the flow, following both filtered DNS and no-model results closely. The over attenuation induced by the plain Smagorinsky model persists as the flow develops and ends up leading to a smaller dissipation peak, in comparison with the other models considered. Furthermore, \modelAcronym's results, when compared to ones obtained via DYN or CvP, are closer to the filtered DNS results from $tV_0/L \approx 6$ onwards, introducing a slight over dampening in the prior period. Similar levels for the peak in $\epsilon_{\mathrm{kin}}$ are recovered for these three models but, only the \modelAcronym-based results recover the dissipation plateau existent in the DNS results after its peak.
}

\review{
\begin{table}[h]
\begin{center}
\begin{tabular}{ c c c } 
 \hline\hline
  & Run time & Computational overhead \\ 
 \hline
 No model &  $t_{\text{ref}}$ & -  \\ 
 SMAG &  1.343$t_{\text{ref}}$ & +34.3\%  \\ 
 CvP &  1.377$t_{\text{ref}}$ & +37.7\%  \\ 
 QSV (only $\tau_{ij}$) &  1.647$t_{\text{ref}}$ & +64.7\%  \\ 
 DYN &  1.725$t_{\text{ref}}$ & +72.5\%  \\ 
 QSV &  1.785$t_{\text{ref}}$ & +78.5\%  \\ 
 \hline\hline
\end{tabular}
\caption{Computational time comparison for the different modeling techniques considered in $96^3$ simulations of a TGV with initial amplitude of $M_t = 0.1$.}
\label{tab:TGVCompCost}
\end{center}
\end{table}
}

\review{The improved ability to solve the transitional flow setup is associated with an increase in computational cost, though. Table \ref{tab:TGVCompCost} compares the computational time necessary to perform a given number of iterations in the simulation of a $96^3$ $M_t = 0.1$ TGV when using the various models considered against the reference time obtained when no model was used. It can be observed that despite leading to over dampening, the Smagorinsky model is the least computationally intensive model studied. Furthermore, the CvP method \citep{Chapelier_JCP_2018} is able to achieve a big improvement in resolution power with little increase in computational overhead. The cost of \modelAcronym~and DYN models are comparable but much more expensive in relation to SMAG and CvP. Ultimately, although the full \modelAcronym~closure is more expensive than the DYN procedure, it is also able to perform shock capturing if needed, as discussed in subsection \ref{subsec:ShockCapturing}. Moreover, if \modelAcronym~is only used to estimate the SFS stress tensor term ($\tau_{ij}$) while neglecting the SFS thermal energy flux ($q_j$), it is shown to be computationally cheaper than DYN. This fact is relevant because it shows that the \modelAcronym~implementation can be made more economical in comparison with the DYN model in incompressible settings. }

\begin{figure}[ht]
\centering
\includegraphics[width=1.\linewidth]{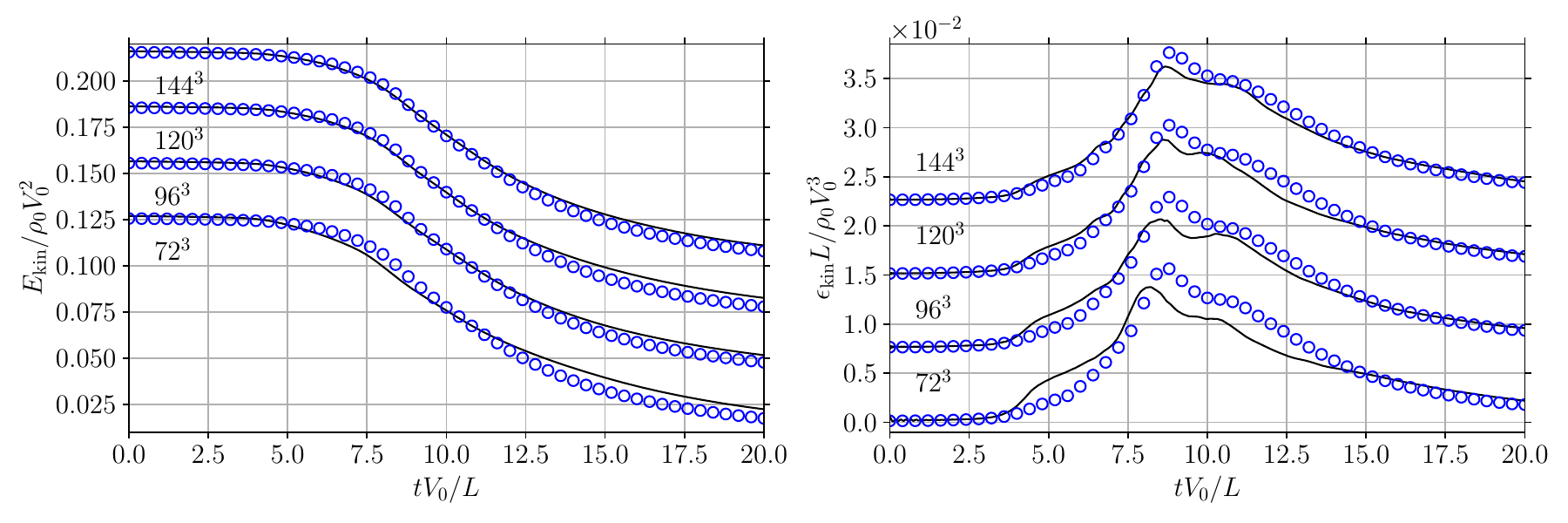}
\put(-250,130){a)}\put(-25,130){b)}
\caption{\review{Grid convergence study for the \modelAcronym~LES of a TGV at $M_t = 0.1$ and $Re = 5000$. Displayed are the evolution of volume-averaged \review{kinetic} energy (a) and its dissipation (b) for $72^3$, $96^3$, $120^3$ and $144^3$ grid points compared against the results acquired from a sharp-spectral filtered DNS up to the same cutoff. Results are vertically shifted for clarity. 
}}\label{fig:TGV_M0-1_TKEepsGrid} 
\end{figure}

A grid convergence study is  also performed on the \modelAcronym-LES of a $M_t = 0.1$ TGV at  $Re = 5000$. Figure \ref{fig:TGV_M0-1_TKEepsGrid} compares the \modelAcronym~results obtained at each grid resolution to the energy and dissipation rate recovered from DNS data spectrally filtered to match the LES grid in question. The results show that each increase in resolution induces an improvement in the predicted time evolution of the simulated volume-averaged energy and its dissipation rate in both delaying the initial breakdown and achieving a dissipation plateau after the peak. 

\subsection{High-speed turbulent channel flow}

The assessment of the \modelAcronym~model in presence of solid boundaries is tackled via compressible turbulent channel flow simulations at supersonic and hypersonic bulk Mach numbers, $M_b = 1.5$ and $6.0$, respectively. Flow parameters are non-dimensionalized using the channel's half-width ($\delta$),  the speed of sound at the wall ($a_w$), the wall temperature ($T_w$) and the bulk density,
\begin{equation}
\rho_b = \langle \rho \rangle_{\Omega},
\end{equation}
\noindent where the bracket $\langle \cdot \rangle_{\Omega}$ represents the volume average over the computational domain. The bulk Reynolds number 
\begin{equation}
\quad Re_b = \dfrac{\rho_b U_b \delta}{\mu_{\mathrm{ref}}},
\label{eq: Re_def} 
\end{equation}
\noindent is also assigned, where $U_b$ is the bulk velocity defined as $U_b = \langle\rho u_1 \rangle_{\Omega}/\rho_b$ and where $\mu_{ref}$ is the reference dynamic viscosity. Additionally, the bulk Mach number can be defined in terms of the aforementioned scales as $M_b = U_b/a_w$ and a power law given by 
\be
\frac{\mu}{\mu_{\mathrm{ref}}} = \left(\frac{T}{T_{\mathrm{ref}}} \right)^n,
\ee
\noindent with $n = 0.76$ is assumed for the dynamic viscosity.

The flow settings are inspired in the work of \citet{ChenS_JFM_2021} who, for a given $M_b$, adjusted the value of $Re_b$ so that the friction-based Reynolds number that accounts for variable density effects \citep{Huang1995},

\begin{equation} \label{eqn:Restar}
Re^*_{\tau} = \frac{\overline{\rho}(\delta)\sqrt{\overline{\tau_w}/\overline{\rho}(\delta)}}{\overline{\mu}(\delta)}\delta,
\end{equation}

\noindent would remain approximately constant for each grid resolution level. Although table \ref{tab:ChannelPar} shows that this is achieved by the current setup, the coarse resolutions considered in this manuscript, chosen to put the turbulence closures performance to test, are not sufficient to recover the DNS-predicted $Re_{\tau}^* \approx 220$ reported in the work of \citet{ChenS_JFM_2021}. As the grid is refined, though, the difference between the values recovered by the current simulation and the DNS gets smaller.

\begin{table}[h]
\centering
\begin{tabular}{ccccccccccc}
\hline \hline

  $M_b$ & $Re_b$ & $L_x \times L_y \times L_z$ & $N_x \times N_y \times N_z$  & $Re_{\tau}$ & $\Delta x^+$ & $\Delta y^+_{\mathrm{min}}$ & $\Delta z^+$ & $Re_{\tau}^*$ \\
\hline                               

 \multirow{3}{*}{1.5}  &  \multirow{3}{*}{5000} & \multirow{3}{*}{$12\delta \times 2\delta \times 4\delta$}& $32 \times 72 \times 16$  & 258& 99& 0.29& 68& 162\\
& & & $64 \times 108 \times 32$ & 300& 57& 0.22& 38& 193 \\
& & & $128 \times 160 \times 64$ & 325& 30& 0.16& 20& 213\\
\hline
 \multirow{3}{*}{6.0}  &  \multirow{3}{*}{20000} & \multirow{3}{*}{$16\delta \times 2\delta \times 4\delta$}& $80 \times 90 \times 40$  & 2270& 460& 2.0& 232& 164 \\
& & & $160 \times 135 \times 80$& 2500& 251& 1.5& 126& 195\\
& & & $320 \times 200 \times 160$& 2640& 132& 1.0& 66& 212\\

\hline \hline
\end{tabular}
\caption{Parameters for the turbulent channel simulations performed in the current work.}
\label{tab:ChannelPar}
\end{table}

\review{High-speed turbulent channel flow calculations were performed using the parameters in table \ref{tab:ChannelPar} with both the \modelAcronym~approach as the SFS turbulence closure and the eddy-viscosity model proposed by \citet{Vreman_PoF_2004}, run with the parameters specified by \citet{ChenS_JFM_2021}.} Results for the transformed mean velocity 
\begin{equation} \label{eqn:trettel}
U^+_{\mathrm{TL}} = \int_0^{\check{u}/u_{\tau}} \left(\frac{\overline{\rho}}{\rho_w}\right)^{1/2} \left[ 1 + \frac{1}{2} \frac{1}{\overline{\rho}} \frac{\partial \overline{\rho}}{\partial y}y - \frac{1}{\overline{\mu}} \frac{\partial \overline{\mu}}{\partial y}y\right]d(\check{u}/u_{\tau}), \quad \mathrm{with} \quad u_{\tau} = \sqrt{\overline{\tau_w}/\overline{\rho_w}},
\end{equation}
 
\noindent plotted against the semi-local wall coordinate \citep{Morkovin_MT_1962,Huang1995}, 

\begin{equation} \label{eqn:ystar}
y^* = \frac{\overline{\rho}(y)u_{\tau}^*}{\overline{\mu}(y)}y, \quad \mathrm{where}\quad u_{\tau}^* = \sqrt{\overline{\tau_w}/\overline{\rho}(y)}, 
\end{equation}

\noindent are used to assess the performance of each model. This velocity transformation was introduced by \citet{Trettel2016} to account for both variable density and heat transfer effects, which are non-negligible in the current setup. 

\begin{figure}[ht]
\centering
\includegraphics[width=1.\linewidth]{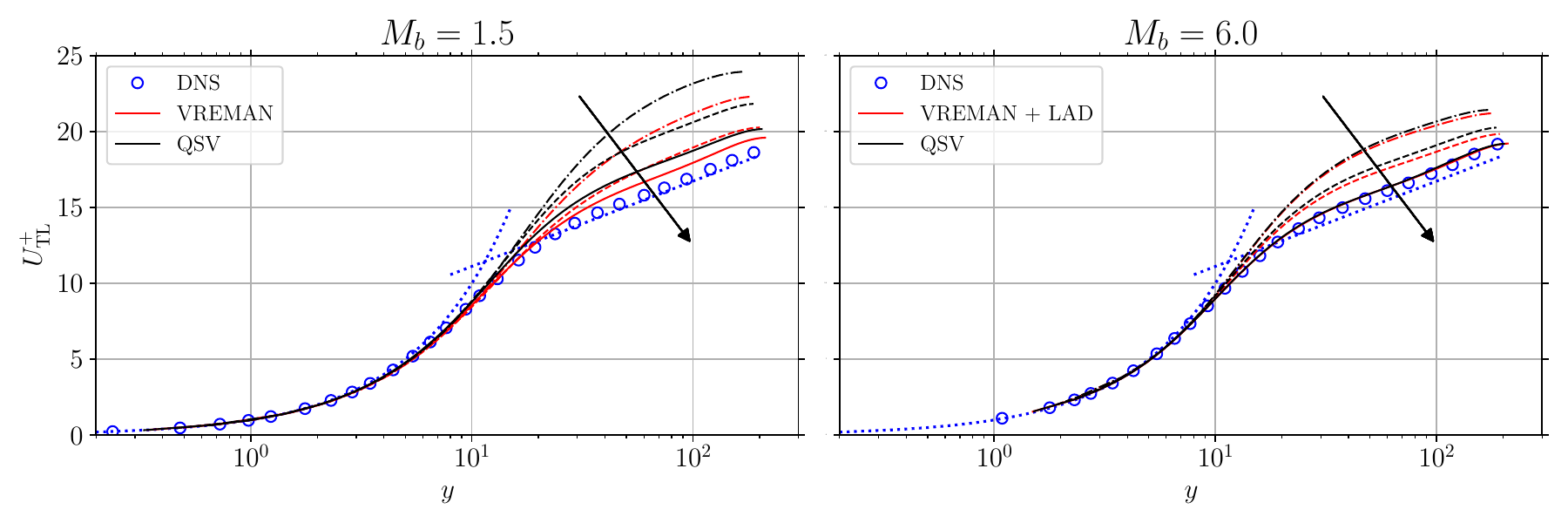}
\caption{\review{TL-transformed \citep{Trettel2016} mean velocity profiles for high-speed channel simulations at different bulk Mach numbers and grid resolutions. An arrow indicates the direction of grid refinement. Different line styles are used for each grid resolution: coarse \sampleline{dashdotted},  intermediate coarse \sampleline{dashed}, fine \sampleline{} (see table \ref{tab:ChannelPar}). The runs were performed using the \citet{Vreman_PoF_2004} and \modelAcronym~methods. The hypersonic $M_b=6.0$ needed the addition of the LAD model to be stabilized. Reference values for the viscous sublayer and the logarithm region are shown in blue dotted lines and DNS data from \citet{ChenS_JFM_2021} is shown in blue symbols.}} 
\label{fig:Channel_TL} 
\end{figure}

Figure \ref{fig:Channel_TL} shows that both models at all grid resolution levels are able to capture the correct viscous sublayer scaling. The performance difference between the models is observed in the logarithmic region, where, at each grid resolution level, the Vreman's model is closer than \modelAcronym~to the reference log-law profile,

\begin{equation}
U^+ = \frac{1}{\kappa} \ln(y^{*}) + C \quad \mathrm{where} \quad C=5.5 \quad \text{and} \quad \kappa = 0.41.
\end{equation}

\noindent and to the DNS results. \review{At $M_b=6.0$, the observed difference between the resolution power of the models considered at each grid resolution is relatively small. At such hypersonic speeds, Vreman's model had to be augmented by the LAD approach to achieve numerical stability. On the other hand, the \modelAcronym~approach is able to solve the setup with the same framework used in all test cases analyzed in this manuscript. At $M_b=1.5$, though, bigger differences between the results achieved by the \modelAcronym~and Vreman closures are observable. Nonetheless, as the grid is refined, the data generated by both models approach the reference DNS results and log-law curves. }

Despite not being the best approach for flow setups with weak compressibility effects, these results show that the \modelAcronym~closure is also applicable to wall-bounded turbulent flows. Therefore, future work could exploit \modelAcronym's ability to perform both shock capturing and turbulence modeling to perform shock/boundary-layer interaction simulations using a single model, for example.

\review{
\subsection{Summary}

In summary, simulations of a subsonic Taylor-Green Vortex (TGV) and turbulent channel flow at $M_b=1.5$ and $6.0$ showcase \modelAcronym's ability to act as a SFS turbulence model. In the TGV test case, \modelAcronym~demonstrated high resolution capability at an increased computational cost, being in the same cost range as the Dynamic procedure. In the turbulent channel test cases, the presence of a non-zero mean shear component led to higher SFS dissipation levels and a decrease in performance compared with Vreman's method. This is particularly true at $M_b=1.5$, where only weak compressibility effects are present. At $M_b=6.0$, the models considered lead to similar results. Ultimately, although the current version of the \modelAcronym~model may not be the best choice for simple wall-bounded turbulent flows up to supersonic speeds, its ability to perform both shock capturing and turbulence modeling simultaneously may prove advantageous in flow setups where shocks and turbulence interact. This is discussed hereafter, in section \ref{sec:QSVboth}. 
}

\review{\section{\modelAcronym's as a unified approach for turbulence modeling and shock capturing} \label{sec:QSVboth}

\subsection{Supersonic Taylor-Green Vortex (TGV)}}

\begin{figure}[ht]
\centering
\includegraphics[width=1.\linewidth]{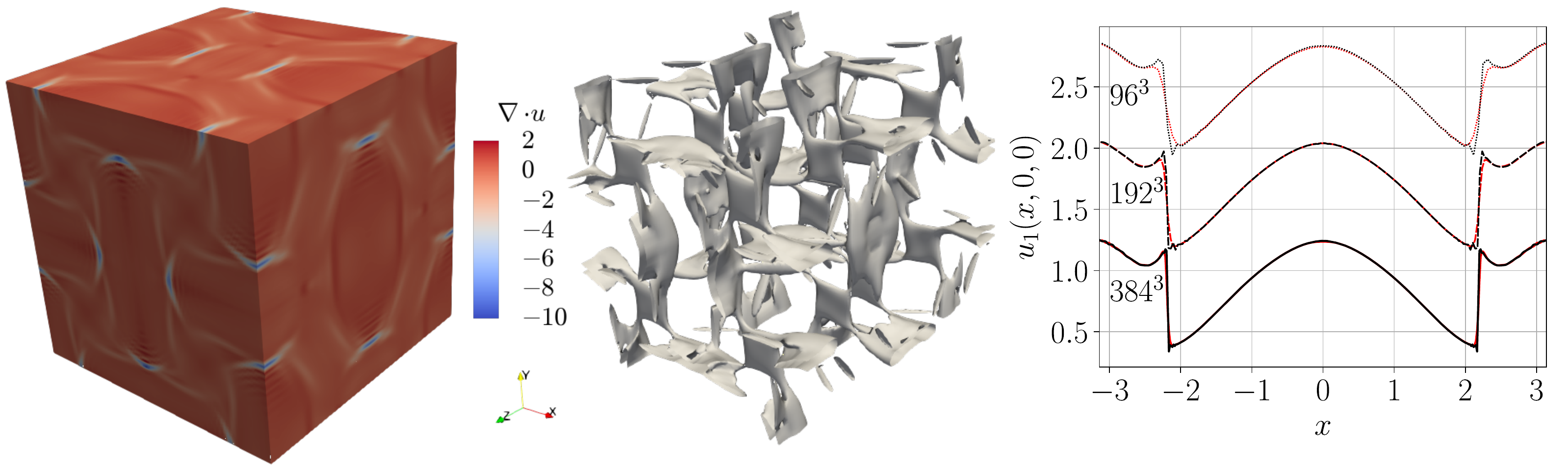}
\put(-460,130){a)}\put(-290,130){b)}
\caption{Visualization of a $96^3$ simulation of a TGV with $M_t = 1.2$ and $Re = 5000$ at time $tV_0/L = 6.75$ displaying divergence field contours (a) and its isosurfaces at $\nabla \cdot u = -0.5$ (b) is shown on the left together with the velocity field at the centerline at different grid resolutions for both \modelAcronym~(black) and LAD augmented Smagorinsky (red), on the right.
}\label{fig:TGV_M1-2_Showoff} 
\end{figure}

\review{After having established the capability of the \modelAcronym~model to act separately as a shock capturing and a SFS turbulence modeling closure, we now test a case exhibiting shock-wave turbulence interaction such as a supersonic TGV flow with initial Mach number $M_t = 1.2$.} At this level, the initial perturbations rapidly induce wave steepening and shocks (see figure \ref{fig:TGV_M1-2_Showoff}) in the solution before the initial vortex breaks down into turbulence and the hydrodynamics start to govern the flow. The objective of this test case is to assess how the different models cope with the presence of both shocks and turbulence in the simultaneously.

Previously, though, the concept of a compressible energy norm will be introduced since there is need to extend the kinetic energy notion, used in incompressible flows to monitor turbulence, into setups in which the compressibility effects are important. With the objective of investigating the possibility of measuring acoustic power transmission directly in convoluted flows, \citet{Myers1991JFM} derived an exact equation governing the energy transported by fluctuations of arbitrary steady base flows. The derived exact compressible energy norm is 

\begin{equation}
E_{\mathrm{c}}(\bf x) = \rho \left[ (H - H_0) - T_0 (s - s_0)\right] - {\bf \rho_0 u_0} \cdot {\bf (u-u_0)}  - (p-p_0),
\end{equation}

\noindent where $H$ is the total enthalpy, $s$ is the entropy and the subscript $_0$ is used to indicate base state quantities, the initial flow quantities in the current case. \review{This is the correct energy norm for compressible flows as it encompasses perturbations up to any order and it accounts for the energy stored in both hydrodynamic and thermodynamic fields.} Additionally, in the case of the TGV, the base flow velocity field is identically zero (${\bf u_0} = 0$) and the relation above can be simplified to

\begin{eqnarray}
E_{\mathrm{c}}(\bf x) &=&  \rho(h - h_0) + \frac{1}{2}  \rho {\bf u^2}  - \rho T_0  (s - s_0) - (p-p_0), \\
\end{eqnarray}

\noindent  Note that, for the case where $M_t = 0.1$, the compressibility effects \review{that lead to energy storage in the enthalpy, entropy or pressure fields} can be neglected and the exact compressible energy norm reduces to the kinetic energy norm.  \review{As in the subsonic TGV case, the exact compressible norm can be volume-averaged, 

\begin{equation}
E_{\mathrm{c}} = \langle \rho(h - h_0) + \frac{1}{2}  \rho {\bf u^2}  - \rho T_0  (s - s_0) - (p-p_0) \rangle,
\end{equation}

\noindent and its dissipation rate can be determined by a time derivative as,

\begin{equation}
\epsilon_{\mathrm{c}} = - \frac{\partial E_{\mathrm{c}}}{\partial t}.
\end{equation}
}

\begin{figure}[ht]
\centering
\includegraphics[width=1.\linewidth]{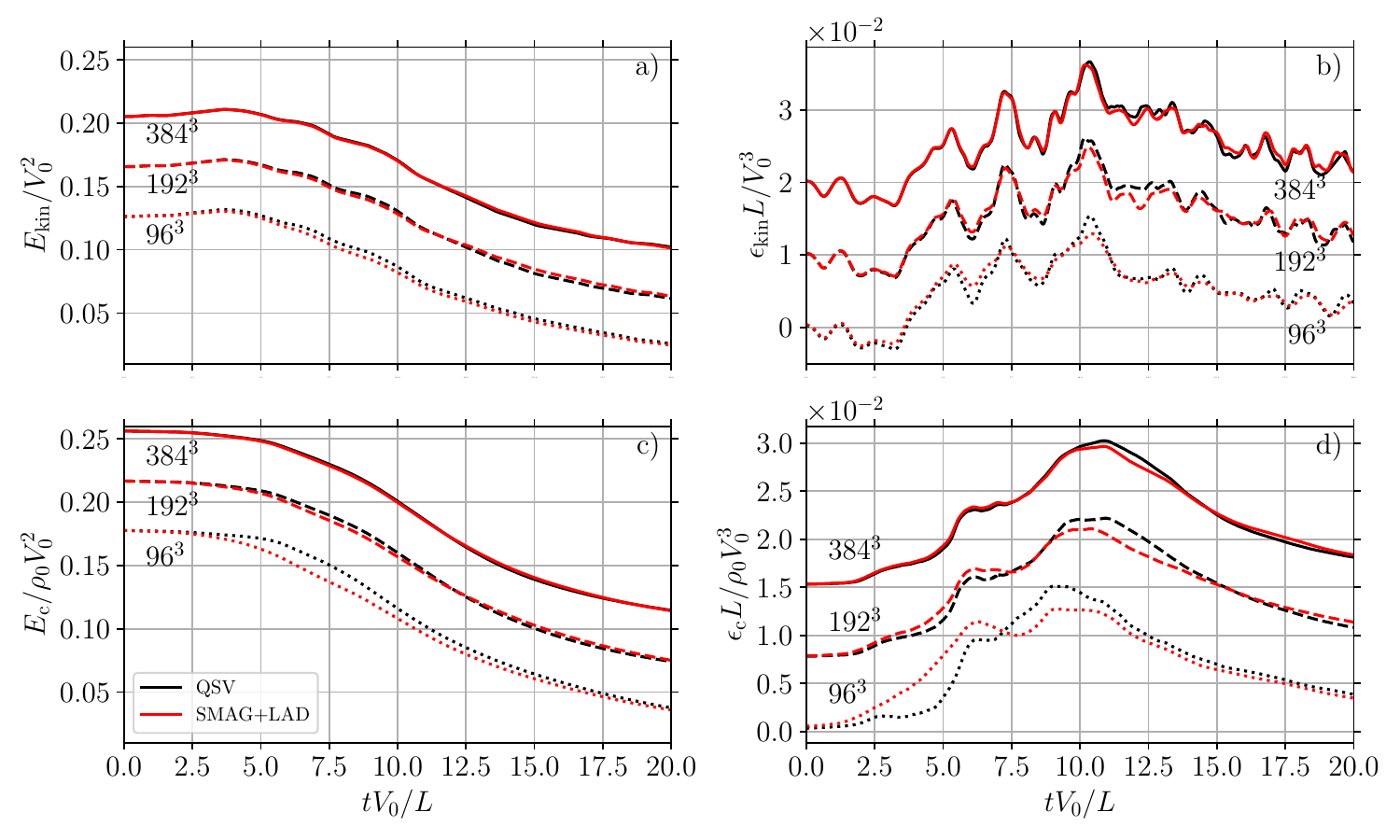}
\caption{\review{Evolution of both volume-averaged kinetic energy (a) and exact compressible energy (c) norms as well as their dissipation rates (b) and (d) in a $M_t= 1.2$ TGV flow at different grid resolution levels. Results in plots (a)-(d) are vertically shifted for clarity.}
}\label{fig:TGV_M1-2_GridConv} 
\end{figure}

\review{Figure \ref{fig:TGV_M1-2_GridConv} gathers the evolution of the two different energy norms, $E_{\mathrm{kin}}$ and $E_{\mathrm{c}}$, as well as their dissipation rate for the TGV with $M_t = 1.2$ and $Re = 5000$. It can observed that the volume-averaged kinetic energy displays a non-monotonic behavior with a non-physical energy growth, (subfigure (a)), and an also fictitious negative dissipation (subfigure (b)). Since there is no actual energy being added to the system and since energy cannot be spontaneously generated, this behavior actually indicates that there is energy exchange between the velocity field and the thermodynamic quantities, which is not accounted in the kinetic energy norm.
Since the TGV with $M_t = 1.2$ displays both compressibility and nonlinear effects, the exact compressible energy norm, $E_{\mathrm{c}}$, is necessary to accurately display the dynamics of this flow. It can be seen that, by using this norm, the expected monotonic energy decay (subfigure (c)), and positiveness of the energy dissipation rate (subfigure (d)), are preserved.} 

\begin{figure}[ht]
\centering
\includegraphics[width=.6\linewidth]{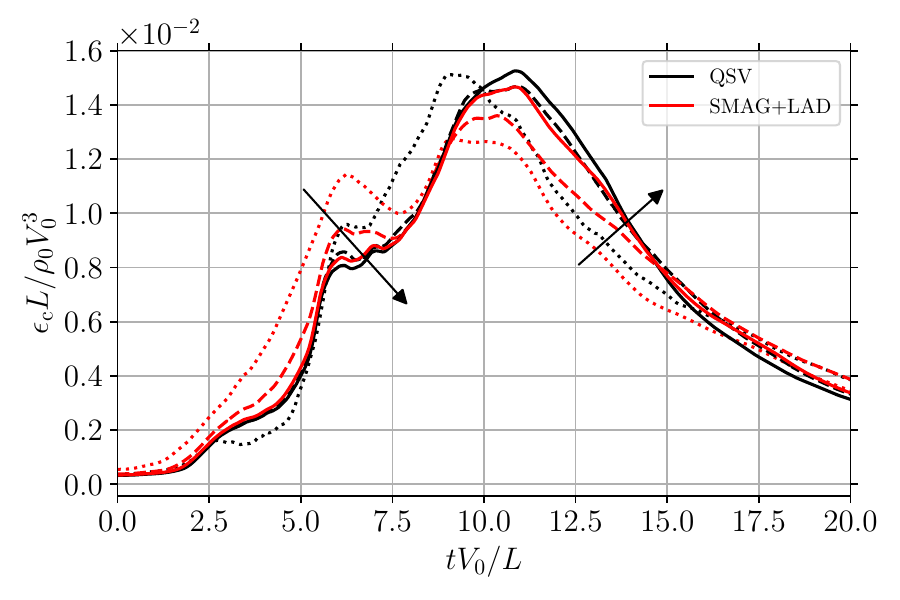}
\caption{\review{The curves obtained for the compressible energy norm dissipation rate, shown in figure \ref{fig:TGV_M1-2_GridConv} (d) are displayed here without the vertical shift to showcase grid convergence, where the arrows point toward the grid refinement direction. }
}\label{fig:TGV_M1-2_GridConv1} 
\end{figure}

\review{Results for the compressible energy norm dissipation rate, shown in figure \ref{fig:TGV_M1-2_GridConv} (d), are repeated in figure \ref{fig:TGV_M1-2_GridConv1} without the vertical shift. From the analyses of such plot, it can be concluded that the results from both the \modelAcronym~and the LAD-aided Smagorinsky simulations converge as the grid is refined. Only slight differences are observed at the $384^3$ grid refinement level, even though the Smagorinsky model is not capable of deactivating its SFS dissipation in the initial phases of the flow, when only large scales are present. Due to the presence of a shock in the flow field, a DNS is not strictly possible. Nonetheless, the convergence of the solutions obtained by different models as the grid is refined supports the use of finest grid considered here as the reference solution.
}

\begin{figure}[ht]
\centering
\includegraphics[width=1.\linewidth]{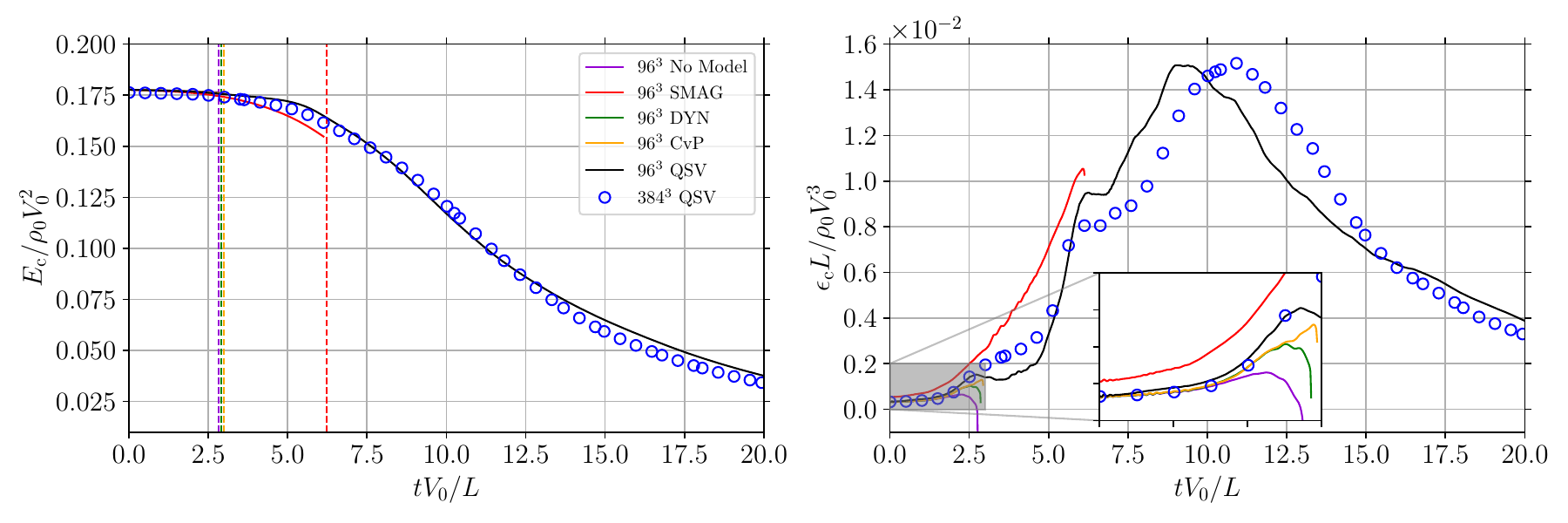}
\put(-420,133){a)} \put(-195,133){b)} 
\\
\includegraphics[width=1.\linewidth]{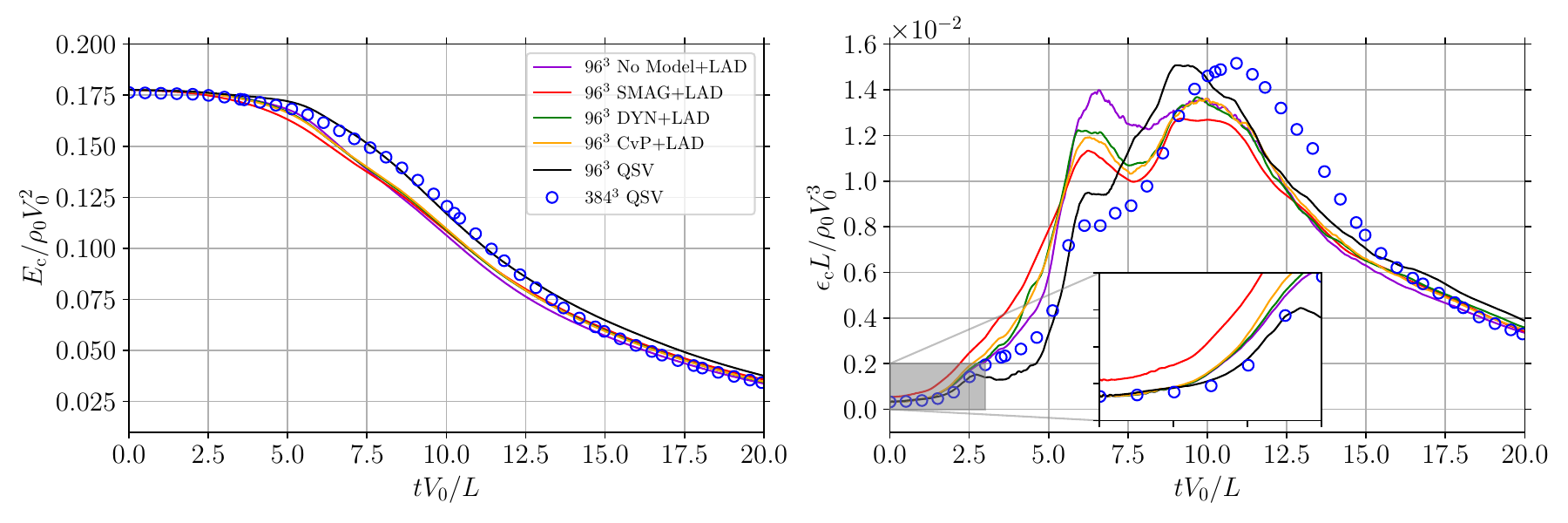}
\put(-420,133){c)} \put(-195,133){d)} 

\caption{\review{Evolution of the volume-averaged exact compressible energy norm, (a) and (c), and its dissipation rate, (b) and (d), in a $M_t = 1.2$ TGV simulation. Results obtained by different sub-filter models using $96^3$ grid points are compared against a reference solution computed using \modelAcronym~model with $384^3$ points. On the top row, (a) and (b), simulations were run without the addition of a shock capturing model, which lead to unstable simulations and their blowup time is shown, in (a), as vertical dashed lines (\sampleline{dashed}). These simulations were then augmented with the LAD shock capturing approach, shown on the bottom row, (c) and (d). }
}\label{fig:TGV_M1-2_TKEeps} 
\end{figure}

\review{The results obtained in the \modelAcronym-based $M_t = 1.2$ TGV simulation with $384^3$ grid points are then used as a reference to compare the performance of the different models considered in a flow setup where both turbulence and shock discontinuities are present simultaneously. First, the results gathered in the upper row of figure \ref{fig:TGV_M1-2_TKEeps} show that, if the current test case is simulated without the inclusion of a shock capturing approach, it leads to numerical instabilities regardless of the turbulence model considered; also when no model is applied. Being the plain Smagorinsky model the most dissipative amongst the ones tested it survives the longest, followed by the CvP, the DYN and the no-model run. The \modelAcronym~model, being able to perform shock capturing, is able to simulate the flow without any numerical instabilities.}

\review{The initially unstable simulations are then augmented by the addition of the LAD shock capturing approach and the results are gathered in the bottom row of figure \ref{fig:TGV_M1-2_TKEeps}. The base formulation of the LAD model \citep{Kawai_JCP_2010} induces the addition of artificial dynamic viscosity ($\mu_{\text{art}}$), which ultimately contributes to the stabilization of the run performed with LAD as the only active model. The results from the LAD model are the most inaccurate and become unstable if $\mu_{\text{art}}$ is deactivated and only the artificial bulk viscosity and artificial conductivity components are active.

Despite achieving numerical stability, all the LAD-aided runs overestimate the dissipation rate around $tV_0/L \approx 6.5$, during the presence of the largest shock discontinuities in the flow (figure \ref{fig:TGV_M1-2_Showoff}), and underestimate the dissipation rate peak, in comparison with the reference results. In comparison with the LAD-aided turbulence models, the \modelAcronym-obtained curve remains closest to the reference results throughout the whole evolution of the flow, introducing less dissipation in the shock-dominated period and  predicting better the magnitude of the dissipation rate peak. Additionally, it can be observed that, in the initial stages of the flow, when only large scales are present \modelAcronym~is the model that is closest to the reference results, showcasing its ability to dynamically modulate the added SFS dissipation magnitude. }

\review{These results ultimately support the claim that the proposed \modelAcronym~model is a genuine unified approach for turbulence modeling and shock capturing. On top of being able to perform each task separately, as discussed in sections \ref{subsec:ShockCapturing} and \ref{sec:QSVturb}, it outperforms the simple addition of separate turbulence and shock capturing models in flow setups where both hydrodynamic turbulence and shock discontinuities are happening simultaneously.}
\\



\section{Conclusion} \label{sec:Conclusion}

A novel technique named the \modelFullName~model (\modelAcronym), was introduced. It is designed to simulate accurately the large scales present in both shock and turbulence dominated flows by exploiting the residual of filter transfer functions to estimate both the amplitude of fluctuations near the grid cutoff and modulate the viscosity magnitude for different wavenumbers. This feature allows for an implementation using only spatial operators, applicable to finite-difference solvers.

The \modelAcronym~mathematical framework is based on an extension of LES closures and a parallel between these and spectral vanishing viscosity (SVV) based models. The 1D Burgers' problem is used to showcase the connection between previous LES models and how they can be \review{understood} as a way of solving shock dominated solutions. Moreover, although the SVV model has been shown to work well when solving equations that allow discontinuous solutions \citep{Tadmor_SIAM_1989,Tadmor_NASA_1990}, a mathematical justification explaining its extension to turbulent flows \citep{Karamanos_JCP_2000,pasquetti2006spectral} was missing. The reasons are presented in the current manuscript in section \ref{sec:IncompressibleLES}: \review{a wavenumber dependent sub-filter flux term is generated by filtering a nonlinear equation. The artificial addition of such dissipation terms done in the aforementioned previous publications, therefore, are similar to performing simulations of the filtered Navier-Stokes equations but failing to explore some of the physical aspects of this consideration.} The SVV kernel, although being inviscid for low wavenumbers, also peaks near the cutoff and can be considered a first order estimate of the exact sub filter flux needed.

Moving forward, the \modelAcronym~model was tested in one-, two- and three-dimensional problems of increasing complexity and it is shown \review{to} perform well in both low-speed and highly compressible flow setups. For example, the same \modelAcronym~framework can be used to solve a Taylor-Green Vortex with both sub and supersonic initial conditions. Moreover, the \modelAcronym~model is flexible, being able to be applied in curved and stretched domains by using grid transformations. The collection of satisfactory results across intrinsically different flow setups supports the claim that the \modelAcronym~method can simultaneously capture shocks and act as a sub-filter turbulent closure.

\review{As a final remark, although the current implementation is aimed at finite difference solvers, it is possible to extend the \modelAcronym~approach to unstructured solvers based on spectral numerics. Due to the opportunity of projecting the solution of each element onto a hierarchical set of orthogonal basis functions, a spectrally based implementation would be able to easily gage the magnitude the energy near the cutoff and would be able to modulate freely the amplitude of the viscosity kernel for different wavenumbers. This fact renders the use of global filtering operations unnecessary and could lead to simpler and more flexible implementations, which will be explored in future research.}


\begin{appendix}

\section{Closure for SFS thermal flux} \label{sec:AppendixThermalFlux}

\begin{figure}[ht]
\centering
\includegraphics[width=1.\linewidth]{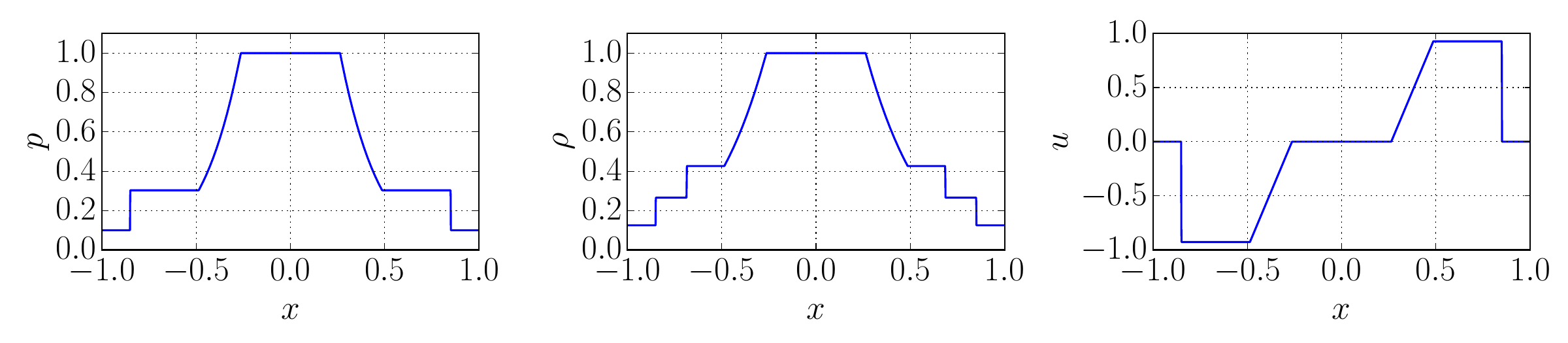}
\caption{Exact solution of the Riemann shock tube \citep{SOD_1978_JCP} problem at dimensionless time $t = 0.2$.
}\label{fig:exactSod} 
\end{figure}

In this section, the Sod shock tube \citep{SOD_1978_JCP}  is used as a canonical test case to further inform the parameters used to model the SFS flux terms in \modelAcronym~approach, specially the thermal flux \eqref{eqn:compqj}. The 1D Euler system of equations augmented by the models for the SFS flux terms discussed in subsection \ref{subsec:1DGasDynamics} are solved using a Fourier based spatial discretization using a 4th-order Runge-Kutta time integration method. To be solvable by a Fourier method, two Riemann interfaces are initialized in the domain $x \in [-1,1]$ with initial conditions,

\begin{equation} 
[\rho,u,p](x,0) = 
\begin{cases}
    [1,0,1],& \text{if } |x| <  0.5\\
     [0.125,0,0.1],              & \text{otherwise},
\end{cases}
\end{equation}

\noindent which leads to an exact solution develops 4 states of constant density separated by symmetric shocks, contact discontinuities and rarefaction waves and it is shown in figure \ref{fig:exactSod} at time $t = 0.2$. This choice was preferred instead of using a direct cosine transform approach because the odd derivatives of the cosine basis functions are not directly represented by the initial orthogonal basis and a new projection operation would be needed. 


\begin{figure}[ht]
\centering
\includegraphics[width=.49\linewidth]{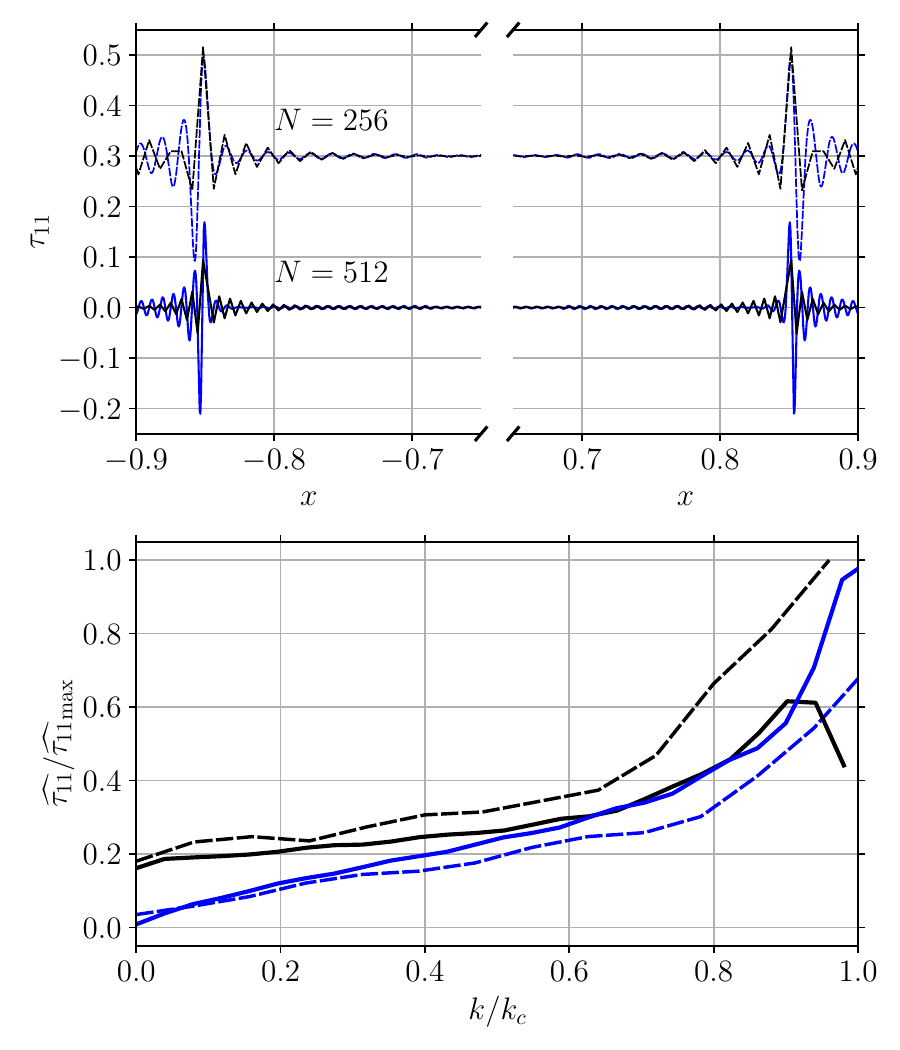}
\includegraphics[width=.49\linewidth]{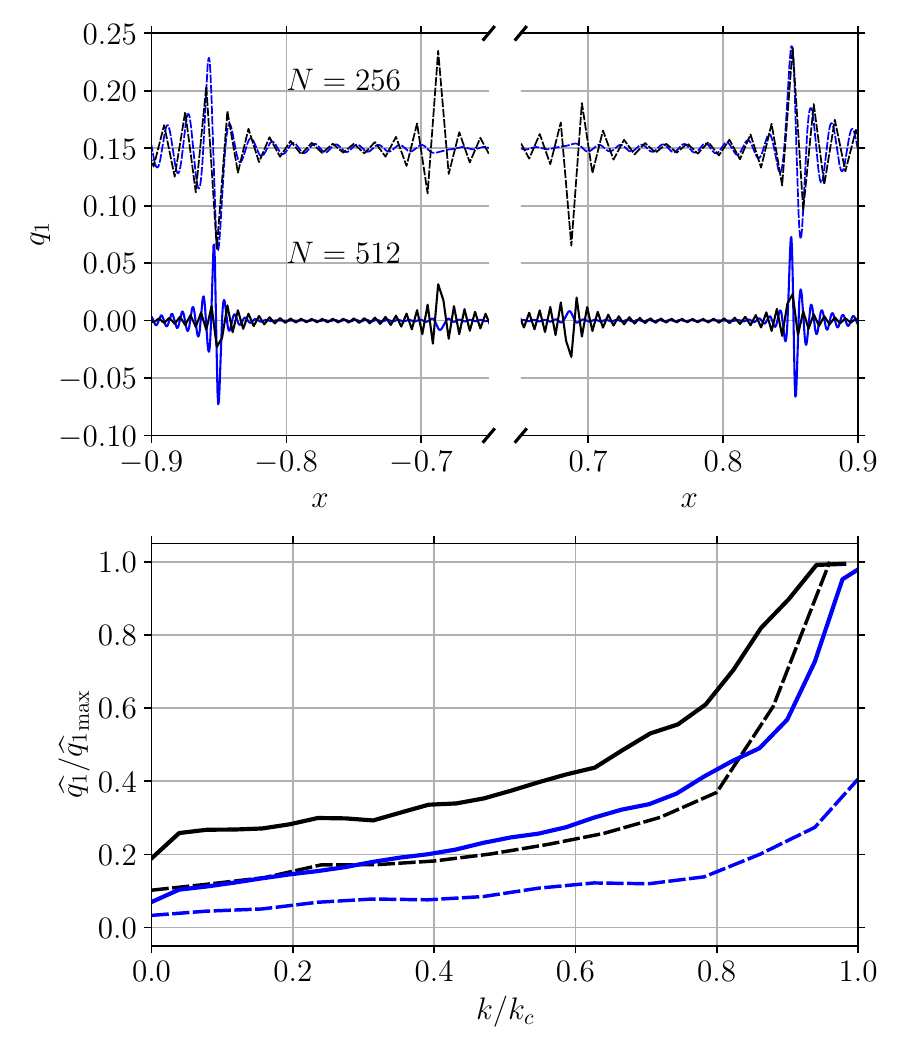}
\caption{\review{Energy flux  to subfilter scales based on the exact solution of the Riemann shock tube \citep{SOD_1978_JCP} problem at  $t = 0.2$ (blue) and values of the SFS fluxes predicted by the proposed QSV closure when fed the exact filtered solution (black). }
}\label{fig:aprioriSod} 
\end{figure}

\review{As discussed at the beginning of subsections \ref{subsec:Burgers} and \ref{subsec:TGV}, an {\it a priori} analysis can reveal the exact energy flux from large to small scales when a sharp spectral filter operation is performed upon a reference solution. Additionally, the filtered reference solution can serve as an input for model in question, which reveals its behavior in both physical and spectral space. The examination of a comparison between the exact flux to sub-filter scales and the one recovered by the model fed by the filtered solution can inform the design of the closure. Such procedure was conducted for the Sod shock tube problem at $t = 0.2$ and its results are shown in figure \ref{fig:aprioriSod}, where the shape of the \modelAcronym's filter modulation transfer function, $\widetilde{G}_{\text{qsv}}$ \eqref{eqn:EffFilterTF}, is further justified due to the observed \emph{plateau-cusp} behavior on both $\hat \tau_{11}$ and $\hat q_1$ quantitites. Additionally, it is possible to observe that the use of the constant $C_q  = 0.8$ in the SFS thermal flux closure \eqref{eqn:compqj} leads to satisfactory matching between exact and modeled dissipation spectra. A slight over attenuation is introduced to help stability during computations of strong shocks while retaining good accuracy.} 

\begin{figure}[ht]
\centering
\includegraphics[width=1.\linewidth]{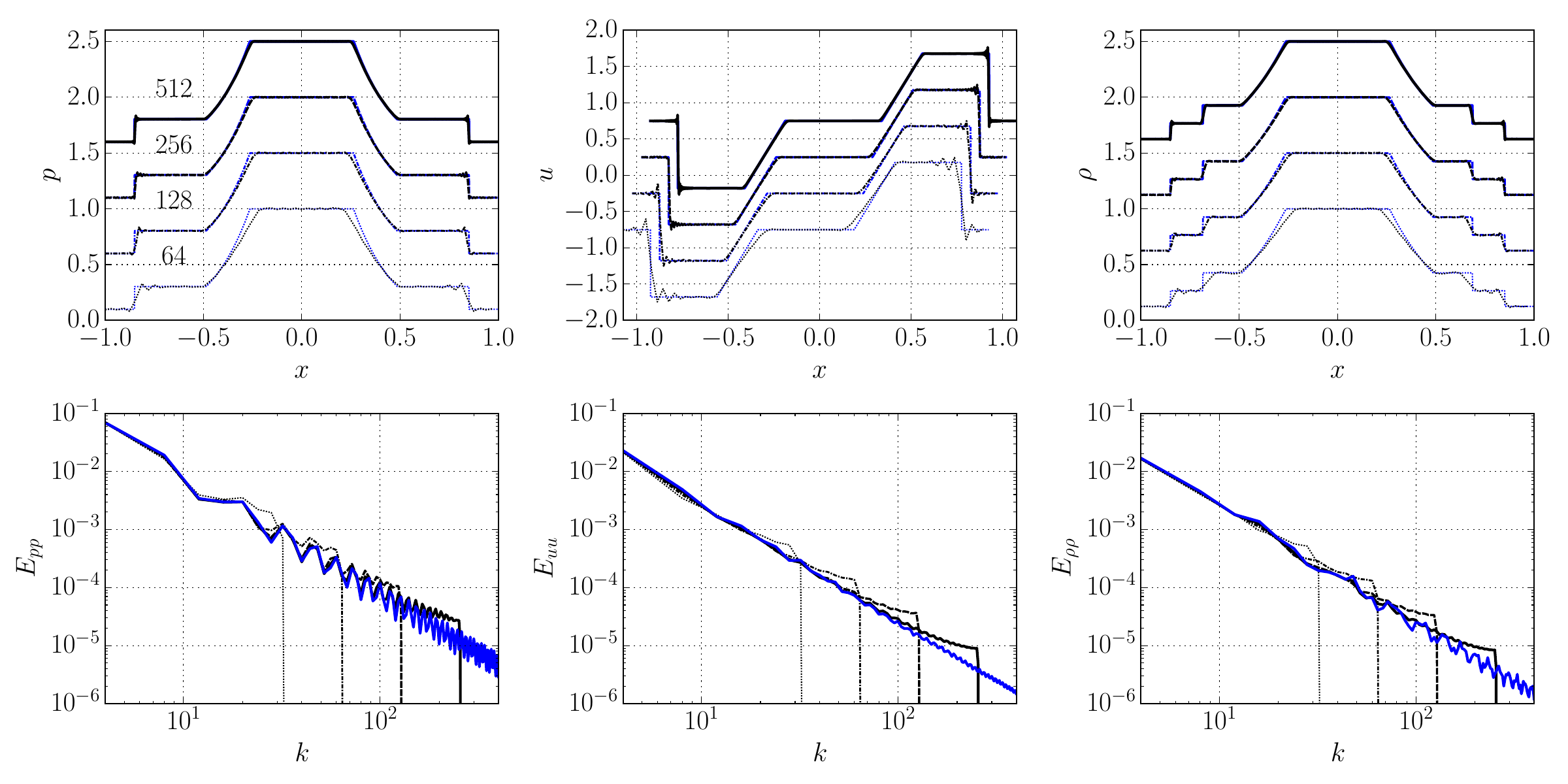}
\caption{\review{ {\it A posteriori} QSV results at different grid resolution levels for the pressure (left), velocity (center) and density (right) fields from a Fourier based numerical simulation of a mirrored 1D Sod shock tube problem are shown in black and are compared against its exact solution, shown in blue, at $t = 0.2$. }
}\label{fig:aposterioriSodPressVel} 
\end{figure}

Following, {\it a posteriori} results of the Sod shock tube problem using the \modelAcronym~closure models in Fourier based pseudospectral simulations are gathered in figure \ref{fig:aposterioriSodPressVel}. The numerical solution for the pressure, velocity and density fields display excellent agreement with the analytical results showing little influence of spurious oscillations even in relatively coarse simulations. In the spectra plots, a small energy accumulation can be observed with little effect on the overall solution quality. 

Although the high frequency oscillations do not pose a stability concern, they can be undesirable in situations where a quiet flow is preferred. In those situations, the numerically implemented scheme can be made more robust in the sense of decreasing the magnitude of these spurious vibrations by explicitly filtering the spectrum in the near-cutoff region of all the conserved variables: density, momentum and internal energy. This explicit filtering step is not needed for the current simple test case, though.

\section{\modelAcronym~in Generalized Curvilinear Coordinates} \label{sec:AppendixB}

Following \citet{Jordan_JCP_1999} and \citet{nagarajan2007leading}, who developed, respectively, the incompressible and compressible LES methodology in generalized curvilinear coordinates, we present the filtered equations and the \modelFullName~method, when subjected to the necessary grid transformations. Assume the existence of a known, invertible mapping between $\mathbf{y}$, the physical cartesian reference frame, and $\mathbf{x}$, the contravariant curvilinear coordinate system,

\begin{eqnarray}
x^i = x^i(y^1,y^2, y^3), \\
y^i = y^i(x^1,x^2, x^3),
\end{eqnarray}

\noindent  where $x^i$ and $y^i$ are the i-th coordinate of each respective system of reference. Following, consider the following curvilinear equivalent of Favre filtering, 

\be
\check f = \frac{\overline{J \rho f}}{\overline{J \rho}}
\ee

\noindent where $J$ is the Jacobian of the transformation which is the determinant of the Jacobi matrix ($J_{ij} = \partial y^i/\partial x^j$), then the filtered governing equations are

\begin{equation}
	\frac{\partial \overline{J \rho}}{\partial t}  + \frac{\partial }{\partial x^j}( \overline{J \rho} \check v^j)= 0,
\end{equation}

\begin{equation} \label{eqn:mom}
	\frac{\partial \overline{J \rho}\check v^i}{\partial t}  + \frac{\partial}{\partial x^j}(\overline{J \rho} \check v^i \check v^j + \overline{J p} g^{ij} - J \check \sigma^{ij} + \overline{J \rho} \tau^{ij})= 
	-\Gamma^i_{qj}(\overline{J \rho} \check v^q \check v^j + \overline{J p} g^{qj} -  J \check \sigma^{qj} + \overline{J \rho} \tau^{qj}),
\end{equation}

\begin{equation}  \label{eqn:ene}
	\frac{\partial \overline{J E}}{\partial t}  + \frac{\partial}{\partial x^j}(\overline{J(E + p)}\check v^j + J \check Q^{j})= \frac{\partial}{\partial x^k}( J \check \sigma^{ij}g_{ik}\check v^k )-  \frac{\partial \overline {J \rho} C_p q^{j} }{\partial x_j},
\end{equation}

\noindent and the subfilter terms are defined as

\begin{equation}
	\tau^{ij} = \widecheck{v^i v^j} - \check v^i \check v^j, \quad \text{and} \quad q^{j} = \widecheck{T v^j} - \check T \check v^j.
\end{equation}

\noindent Moreover, the tensors responsible for mapping a curvilinear physical space into an euclidean reference space are the covariant and contravariant metric tensors $g_{ij} = \frac{\partial y^i \partial y^j}{\partial x^k \partial x^k}$, $g^{ij} = \frac{\partial x^k \partial x^k}{\partial y^i \partial y^j}$, respectively, and the Christoffel symbol of the second kind, $\Gamma^i_{qj} = \frac{\partial x^i}{\partial y^l}  \frac{\partial^2 y^l}{\partial x^q \partial x^j}$. In the derivation of these equations, the metric tensors and Christoffel symbols are assumed to be varying slowly over the spatial support of the filter kernel, therefore leading to no additional sub filter flux terms. 

In the curvilinear frame of reference the total energy, the viscous stress tensor and the heat flux vector are described by slightly modified relations described below:

\begin{equation} \label{eqn:totEnergy}
\frac{\overline{J p}}{\gamma - 1}  = \overline{J E} -  \frac{1}{2}  \overline {J \rho} g_{ij} \check v^i \check v^j -  \frac{1}{2}  \overline {J \rho} g_{ij} \tau^{ij},
\end{equation}

\begin{equation}
	\check \sigma^{ij} = \mu \left(g^{jk} \frac{\partial \check v^i}{\partial x^k} + g^{ik} \frac{\partial \check v^j}{\partial x^k} - \frac{2}{3} g^{ij}\frac{\partial \check v^k}{\partial x^k} \right), 
\end{equation}

\begin{equation}
	\check Q^j = -k  g^{ij} \frac{\partial \check T}{\partial x^i}.
\end{equation}

Ultimately, the proposed closures for $\tau^{ij}$ and $q^{j}$ when applied to generalized curvilinear coordinates are 
\review{
\begin{eqnarray} \label{eqn:fullclosure_tau_curv}
\tau^{ij} &=& - C_{\tau^{ij}} \frac{1}{2} \left( g^{jk} \mathcal{D}^{ik}\frac{\partial \ddot v^i}{\partial x^k} + g^{ik} \mathcal{D}^{jk}\frac{\partial \ddot v^j}{\partial x^k}   \right) ,\\  
 \label{eqn:fullclosure_q_curv}
  q^{j} &=& - C_q  g^{jk} \mathcal{D}^{kk}\frac{\partial \ddot T}{\partial x^k}. 
\end{eqnarray}

\noindent Here, the double dot superscript indicates the filter modulated quantities, defined in curvilinear coordinates as
 \be
 \frac{\partial \ddot v^i}{\partial x^k} = \frac{\partial \check v^i}{\partial x^k}* \left(1 - \widetilde{G}_{\text{qsv}}\right),
 \ee
 \be
 \frac{\partial \ddot T}{\partial x^k} =  \frac{\partial \check T}{\partial x^k}* \left(1 - \widetilde{G}_{\text{qsv}}\right),
 \ee

\noindent where $\widetilde{G}_{\text{qsv}}$ \eqref{eqn:EffFilterTF} is \modelAcronym's filter modulation transfer function and $\mathcal{D}^{ij} = \upsilon^i(\check v^j)  \ell^j$ is the dissipation magnitude tensor comprised by a length scale $\ell^j = \overline \Delta^j$, the computational grid spacing in each direction, defined by the initial mapping from physical space to a reference space where the equations are solved, and the sub-filter velocity scale

\be
\upsilon^i(\check v^j) = \sqrt{ \frac{2 E^i_{k_c}\left(\check v^j \right)}{\overline \Delta^i}}.
\ee
}

\end{appendix}

\section*{Acknowledgments}
Victor Sousa and Carlo Scalo acknowledge the computational support of the Rosen Center for Advanced Computing (RCAC) at Purdue and of the U.S. Air Force Research Laboratory (AFRL) DoD Supercomputing Resource Center (DSRC), via allocation under the subproject AFOSR43032009. This project was funded by the Air Force Office of Scientific Research (AFOSR) grant FA9550-16-1-0209, the AFOSR YIP (FA9550-18-271-0292), the Office of Naval Research YIP (N000142012662) as well as the ONR Grant No. N00014-21-1-2475. Victor Sousa also acknowledges the support of the Lynn Fellowship administered by the interdisciplinary Computational Science and Engineering (CS\&E) graduate program at Purdue University. 

\bibliography{./references.bib}

\end{document}